# A novel geometric predictive algorithm for assessing Compressive Elastic Modulus in MEX additive processes, based on part nonlinearities and layers stiffness, validated with PETG and PLA materials

Jorge Manuel Mercado–Colmenero[12], Cristina Martin–Doñate[12*]

[1]Department of Engineering Graphics Design and Projects. University of Jaen. Spain

[2] Research Group INGDISIG Jaen. University of Jaen. Spain

*Correspondence: cdonate@ujaen.es

Campus Las Lagunillas, s/n. Building A3-210 23071 Jaen (Spain) Phone: +34 953212821, Fax: +34 953212334

## Abstract

MEX (Material Extrusion) is an intrusive technological process that inherently induces alterations in the elastic and mechanical parameters of plastic materials. Manufacturers provide initial mechanical parameters for plastic filaments, which undergo modifications during MEX manufacturing, influenced by intrinsic manufacturing factors such as temperature and pressure changes, as well as geometric and technological parameters of the 3D additive process. These factors, compounded by the inherent geometric nonlinearities in plastic components, directly impact the post-manufacture mechanical and elastic properties of the material. Presently, material characterization in MEX manufacturing relies on manual experimental testing, necessitating new tests for any variation in manufacturing parameters. In this scenario of mechanical uncertainty, rigorously validating component behavior involves costly experimental trials. Intending to solve the problems of MEX components manufacturing, the paper presents an innovative methodology based on the use of a new predictive algorithm created by the researchers capable of obtaining the elastic modulus of a plastic material manufactured with MEX and its mechanical behaviour in the elastic zone under compressive loads. The predictive algorithm only needs as input the compressive elastic modulus of the isotropic plastic material filament and the manufacturing parameters of the MEX process. The smart developed algorithm calculates the stiffness of each layer considering the number of holes in the projected area. The innovative predictive algorithm has been experimentally and numerically validated using PETG (Polyethylene Terephthalate Glycol) material and PLA (Polylactic Acid) on test specimens and on a case study of variable topology. The results show deviations from [0.2% – 4.3%] for PETG and [0.4%] for PLA concerning the experimental tests and [1.1%-13.5%], to the numerical analyses. In this line, the presented algorithm greatly improves the results obtained by the simulation software since this software currently can not consider the geometric and technological parameters associated with the 3D manufacturing process of the component. The predictive algorithm is valid for each print pattern and manufacturing direction. The new algorithm improves the existing state of the art significantly since this algorithm extends its utility to most plastic polymer materials suitable for MEX 3D printing, provided that the mechanical and elastic properties of the filament are known. Its versatility extends to complex component geometries subjected to uniaxial compression loads, eliminating the need for mechanical analysis software or expensive experimental validations.

## 1. Introduction

In recent years, additive manufacturing technology has streamlined the design process for industrial products [1,2]. The use of fast and affordable models, the manufacture of complex designs, and the freedom to conceive innovative and sustainable geometries have made processes such as material extrusion (MEX), initially created for model making, stand out due to its versatility in applications with high mechanical requirements and reduced budgets [3-5].

In a very concise way, the MEX process uses a thermoplastic filament that changes from solid to melting state after passing through a set of resistances in an extrusion head [6]. A nozzle located in the head deposits the molten material in layers, following a predefined pattern established by the device's manufacturing software [7,8]. Despite the multiple advantages of the MEX process, it still presents several technological problems, which causes the manufactured parts to reduce their mechanical capacity in comparison with traditional processes such as injection molding [9-11]. The MEX technology requires the establishment of a series of manufacturing parameters related to the topology of the designed part [12-14]. Identifying the appropriate combination of manufacturing parameters that efficiently maximizes the mechanical properties is a complex task [15-17]. Additionally, and for demanding mechanical stiffness requirements, a complete filling with densities of 100% is required [18]. Although it has approaches that maximize density, the manufactured component exhibits a strong anisotropy, which means that the structural behavior of the part differs significantly from the filament's initial properties. Accordingly, one of the main challenges of the MEX process is limiting the level of uncertainty in the mechanical response of the manufactured parts [19].

The appearance of pores or internal voids is considered a manufacturing defect. The size, location, and density of the pores influence the visual quality [20], the dimensional precision [21], and the mechanical properties [22-26] of the printed part causing structural failures, such as cracks [27], tears [28], fractures [29], delaminations [30] and reducing parameters such as maximum tensile strength (UTS), bending, and impact [31,32]. Numerous research studies have focused on analyzing the mechanical behavior of parts manufactured by MEX, optimizing the process parameters with experimental approaches in an effort to better understand the structural behavior of the printed parts. These models ultimately seek an improvement in the prediction of the mechanical properties of the component, as well as in the optimization of the process parameters. AliSajjadi et al [33] studied the influence of the raster angle on the HIPS material and its response to fracture. The results show that the layers with a raster angle [45°/45°] have the highest strength. Fisher et al [34] studied the properties of ABS Terluran GP 35 with mechanical properties close to the specifications of injection molded parts. By varying the diameter of the nozzle and the layer height, they improved the elastic limit from 55 to 75% compared to the injection process. Vanaeai et al [35] experimentally studied the influence of various process parameters on the thermal and mechanical properties of parts made with PLA material and MEX technology, concluding that the extruder temperature has the greatest influence. Frunzaverde et al [36] carried out a set of tensile tests with the PLA material, varying the layer height and the color of the material. The experimental results showed that the color of the filament is a factor of great influence on the dimensional precision as well as on the tensile strength. Cerdá Avila et al [37] studied the structural behavior of parts made of PLA material by varying the process parameters. The results of the experimental tests showed that the parameters related to the filling and the part orientation were the most influential.

Some research papers have used numerical models to analyze and evaluate the mechanical performance of parts manufactured by MEX. Sammy et al. [38] presented a study focused on the simulation of the MEX manufacturing process for a crystalline PP material using different fill patterns to visualize and predict internal thermal residual stresses and the resulting deformation as a function of printing temperature. Khanafer et al [39] developed a three-dimensional computational model capable of analyzing the transient in the heat transfer problem as well as the behavior of adhesion between layers that affects the mechanical behavior of the part. Unfortunately, the numerical simulation software currently does not have enough information to reliably validate the anisotropic behaviour of the component because it does not take into account the voids caused by the manufacturing process.

Because experimental and numerical models cannot accurately predict the mechanical behavior of MEX-manufactured parts, there is uncertainty regarding their mechanical properties when they are operated. The lack of predictive analytical models has restricted the use of additive manufacturing processes for end-use parts, hindering the evolution of design for additive manufacturing (DfAM) techniques. It is, therefore, necessary to arrange models that can take into account the anisotropic nature of the additive part as well as the internal voids being able to provide a structural behavior understanding. In this line, Garzon et al. [40] proposed an analytical method which seeks to establish a relationship between the porosity and the mechanical properties of the component, linking the Young's modulus and the UTS with the void density and the layer height. The results conclude that a reduction in the height of the layer from 0.3 mm to 0.1 mm decreases the void density by 97%, increasing the Young's modulus and the UTS by 33%. Unfortunately, the research results are presented at a theoretical level without experimental verification in real work conditions. Croccolo et al [41] presented an analytical model capable of predicting the elastic modulus and mechanical stiffness of an additive manufactured plastic material based on the parameters filling pattern angle, layer height, and number of contours. Garg et al [42] used a hybrid genetic algorithm to model the MEX process. The methodology combines genetic programming (GP) with least squares (MS) models using as input parameters the layer height, building orientation, and raster angle obtaining the compressive strength as an output parameter. Casavola et al [43] used the classical laminated theory to analyze the mechanical behavior of the samples manufactured by MEX extrusion. Although the model predicted the elastic modulus for PLA and ABS materials, parameters such as variability in the manufacturing direction of the part were not taken into account. Costa et al [44] presented an analytical solution to analyze the phenomenon of transient heat conduction during the deposition of the plastic filament. Panda et al [45] presented three numerical models to analyze the mechanical properties in cell structures fabricated with MEX tecnology. They used neural networks, genetic algorithms, and surface regression. The results suggest that the neural networks were the ones with the best predictive fit for the case of cellular structures. Christensen et al [46] proposed a model for materials with a porous structure to understand the relationship between porosity and the mechanical properties of materials. According to Garbozzy et al [47], porous materials present a filtration threshold below which there is no porosity. As this threshold is approached and reached, the elastic properties of the material begin to depend on the small remaining voids. Vidakis et al [48, 49] analyze the surface quality of the components manufactured through this additive manufacturing process, focusing on parameters such as roughness, dimensional precision and smoothness. However, the predictive models generated require experimental analysis and manufacturing of a high number of components.

As reflected in the state of the art, although significant advances have been made in this area, so far, no predictive model has been able to develop a solid and reliable predictive model that allows safely approaching the design and production of plastic parts manufactured with the MEX process validating its mechanical response under compressive stress. To solve this problem, the authors, present a novel predictive algorithm that, unlike existing models, is capable of obtaining the elastic behavior of parts manufactured using the MEX process subject to compressive loads, solely from the data on the properties of the material plastic filament and the configuration of the geometrical and technological parameters for the MEX manufacturing process. In a comprehensive and analytically parameterized approach, this methodology facilitates the determination of elastic properties of materials taking to account possible modifications to the geometric and technological parameters within MEX process. Significantly, this obviates the necessity to produce multiple components and conduct numerous tests for each alteration in process configuration MEX. The predictive algorithm presented in this paper is capable of determining the Young's modulus and the part mechanical response in the elastic region of the plastic material under pure uniaxial compression service conditions. Additionally, the newly developed algorithm, unlike previous models, predicts the mechanical behavior of the part taking into account the type of 3D printing infill pattern used in the manufacture of the plastic part and the printing direction in four different 3D printing infill patterns (Concentric, Zig – Zag, Lines 0º/90º and Lines –45º/45º), each of them applied in the main 3D printing directions (Z–axis and X/Y–axis). In this way, it is possible to safely approach the design of a part manufactured with a MEX process without requiring expensive experimental tests or numerical simulations that do not take into account the anisotropy or the real strength of the designed component. The novel algorithm developed by the authors is valid for most polymer plastic material, for its mechanical behavior and elastic - linear regime, and any component with nonlinear geometry.

## 2. Materials and methods

### 2.1. Predictive structural modeling

In order to carry out a characterization of the mechanical and elastic behavior of a plastic material additively manufactured with MEX, the proposed methodology presents a predictive algorithm that allows obtaining the Young's modulus and nominal displacements of a geometry with variable topology within the material elastic region under pure uniaxial compression conditions.

For the development of the predictive algorithm, a series of previous premises have been taken into account:

- The elastic and mechanical behavior of the plastic material filament is considered orthotropic. That is, its elastic and mechanical parameters are defined for its longitudinal direction (Z–axis) and transversal direction (X/Y–axis), respectively (see Fig. 1). Likewise, it is assumed that these properties are not affected by the thermal hysteresis cycle to which the plastic material is subjected during its additive manufacturing process.

- Once it has been submitted to the MEX additive manufacturing process, the section of the plastic filament is considered a slotted hole with dimensions Ø 0.15 mm x 0.15 mm (see Fig. 1).
- The links established between contiguous layers of plastic material are considered welds or rigid joints. Consequently, the geometry resulting from the additive manufacturing process is considered a uniform rigid solid, in which the transversal voids generated during the additive manufacturing process are determined and evaluated (see Fig. 2).
- The geometric parameters of the yarn filament are considered constant throughout the additive manufacturing process. These, in turn, are defined by the parameters that configure the process: line width, layer thickness, and wall thickness (see Fig. 1).
- The behavior of the plastic material is considered linear and only its elastic regime is modeled analytically. Likewise, from the topological analysis, the geometric non-linearities on the manufactured parts can be considered and evaluated. The stress state to which the geometries are subjected is considered static, stationary, and linear.

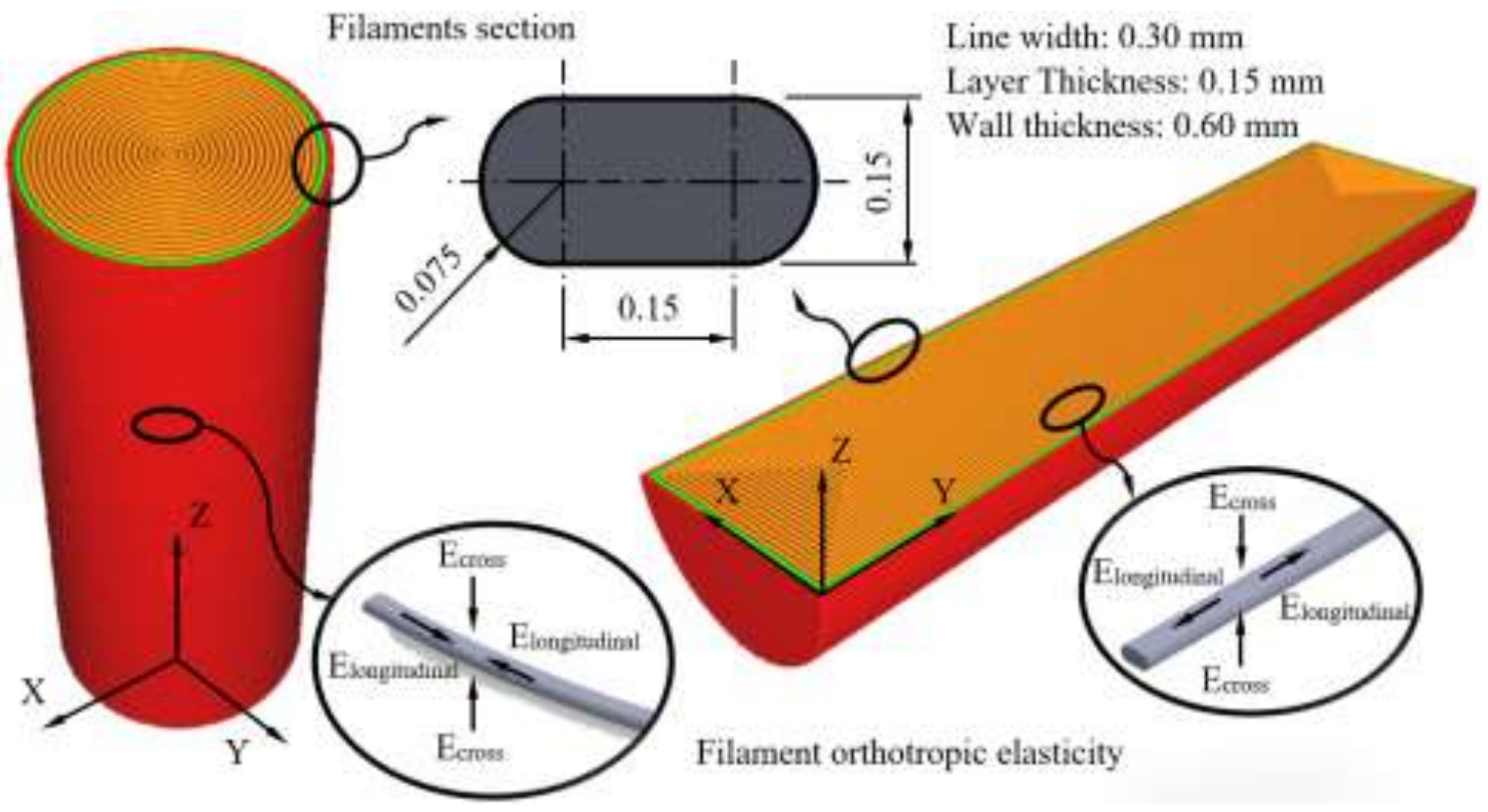

**Fig. 1** Graphical representation of the hypotheses established for the predictive algorithm

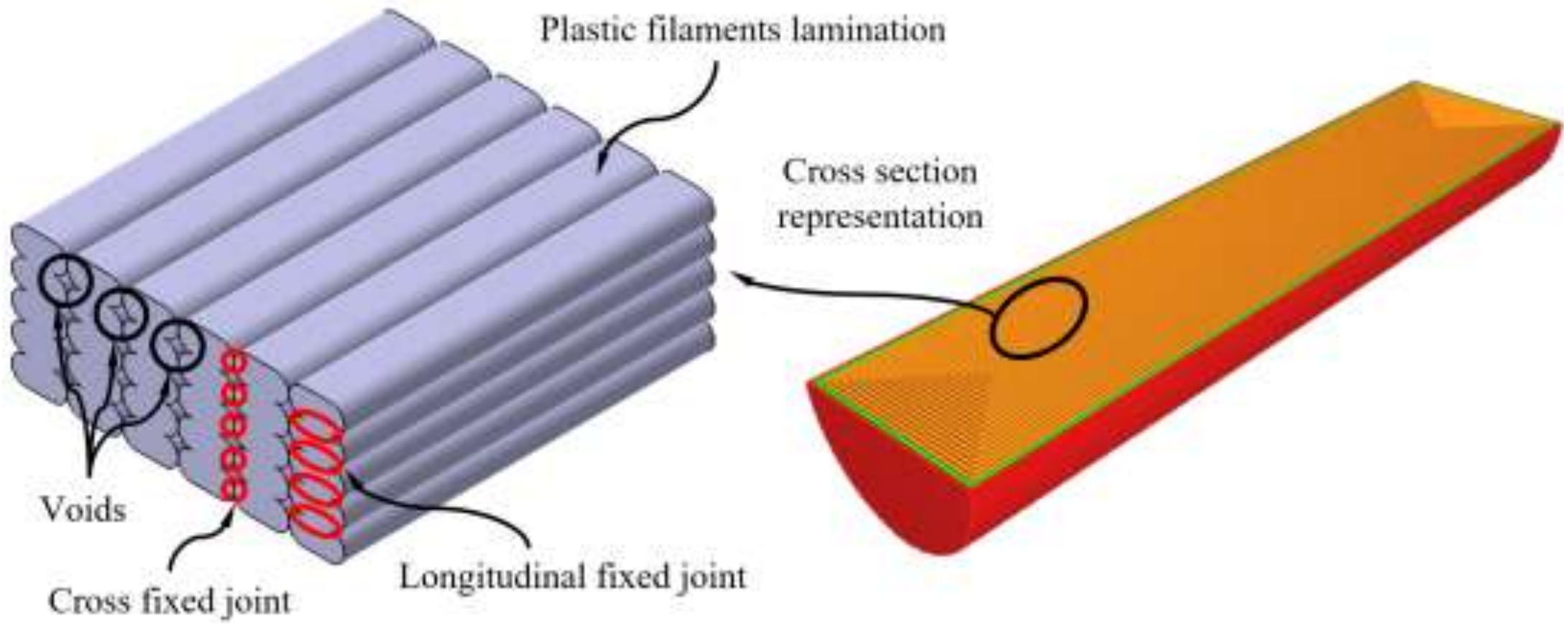

**Fig. 2** Graphic definition of the joints between plastic filaments and voids generated by the manufacturing process

The objective of the developed predictive algorithm is to perform a predictive characterization of the elastic behavior of geometries manufactured by MEX, based on the properties of the plastic material filament (see Table 1) and the configuration defined for its additive manufacturing (see Table 2).

**Table. 1** Properties of the plastic filament

| Properties | Units |
|---|---|
| Material density | g/cm$^3$ |
| Printing temperature | ºC |
| Hot pad temperature | ºC |
| Glass transition temperature | ºC |
| Softening temperature | ºC |
| Filament diameter | mm |
| Cross Young's modulus – $E_{cross}$ – X/Y-Axis | MPa |
| Longitudinal Young's modulus– $E_{longitudinal}$ – Z-Axis | MPa |

**Table. 2** Defined geometrical and technological parameters for the 3D printing manufacturing process

| Variable | Units |
|---|---|
| Line width | mm |
| Layer thickness | mm |
| Wall thickness | mm |
| Wall line count | – |
| Nozzle size | mm |
| Infill density | % |
| Printing speed | mm/s |
| Wall pattern | – |

Assuming that the geometries will work in the elastic zone of the plastic material, the compressive stress – nominal strain relationship can be defined for an orthotropic material, according to the Lamé – Hooke law of elasticity (see Eq. 1). Given that the load condition is pure uniaxial compression, the compressive stress-nominal strain relationship for an orthotropic material can be simplified for a single spatial dimension, as shown described in Eq. 1.

$$\begin{pmatrix} \varepsilon_{xx} \\ \varepsilon_{yy} \\ \varepsilon_{zz} \\ \varepsilon_{xy} \\ \varepsilon_{xz} \\ \varepsilon_{yz} \end{pmatrix} = \begin{bmatrix} \frac{1}{E_x} & -\frac{\upsilon_{yx}}{E_y} & -\frac{\upsilon_{yx}}{E_z} & & & \\ -\frac{\upsilon_{xy}}{E_x} & \frac{1}{E_y} & -\frac{\upsilon_{zy}}{E_z} & & & \\ -\frac{\upsilon_{xz}}{E_x} & -\frac{\upsilon_{yz}}{E_y} & \frac{1}{E_z} & & & \\ & & & \frac{\upsilon_{yx}}{E_y} & 0 & 0 \\ & & & 0 & \frac{\upsilon_{yx}}{E_y} & 0 \\ & & & 0 & 0 & \frac{\upsilon_{yx}}{E_y} \end{bmatrix} \cdot \begin{pmatrix} \sigma_{xx} \\ \sigma_{yy} \\ \sigma_{zz} \\ \sigma_{xy} \\ \sigma_{xz} \\ \sigma_{yz} \end{pmatrix} \rightarrow \varepsilon_{zz} = \frac{\sigma_{zz}}{E_z} \tag{1}$$

Where $\varepsilon_{zz}$ [mm/mm] represents the resulting nominal strain in the longitudinal direction of application of the load conditions, $\sigma_{zz}$ [MPa] represents the stress resulting in the longitudinal direction of application of the load conditions, and $E_z$ [MPa] the compressive Young's Modulus of the filament of plastic material in the longitudinal direction of load application. As indicated in Table 1, the plastic material filament has an orthotropic elastic characterization. Therefore, the elastic behavior of the part geometry varies depending on the main print direction defined during its 3D additive manufacturing process. As shown in Fig. 1, for the Z–axis printing direction the compressive Young's Modulus corresponding to the longitudinal direction of the load scenario application is $E_{cross}$ [MPa] (see Table 1). And, secondly, for the X/Y axis printing direction, the compressive Young's Modulus corresponding to the longitudinal direction of the load scenario application is $E_{longitudinal}$ [MPa] (see Table 1). From Eq. 1, and given that the defined service condition is pure uniaxial compression, the resulting stress $\sigma_{zz}$ [MPa] obtained on the elastic-linear elements can be defined as the ratio between the pure uniaxial compression force applied and their corresponding cross-sectional area (see Eq. 2). Similarly, the field of nominal strains $\varepsilon_{zz}$ [mm/mm], for a pure uniaxial compression load, is defined as the relationship between the field of displacements obtained on the geometries and their total length (see Eq. 2).

$$\sigma_{zz}\,(A_{cross}) = \frac{F_z}{A_{cross}}\;;\;\varepsilon_{zz} = \frac{\Delta L_z}{L_z}$$

(2)

Where $F_z$ [N] represents the pure uniaxial compression force applied on the geometry, $A_{cross}$ [mm$^2$] represents the cross-sectional area, $\Delta L_z$ [mm] represents the field of displacements, and $L_z$ [mm] represents their total longitudinal dimension. Substituting Eq. 2 in Eq. 1, the relationship between the pure uniaxial compression force and the field of resulting displacements, derived from the Lamé – Hooke law of elasticity (see Eq. 3), is defined.

$$\varepsilon_{zz} = \frac{\sigma_{zz}}{E_z} \rightarrow \frac{\Delta L_z}{L_z} = \frac{F_z}{E_z \cdot A_{cross}} \rightarrow \Delta L_z = \frac{F_z \cdot L_z}{E_z \cdot A_{cross}}$$

(3)

In this way, the uniaxial pure compression force to which the geometry is subjected is related to the displacement field generated by the compressive Young's Modulus and its geometric features, height, and cross-sectional area. However, as occurs with the plastic parts, the cross-sectional area is not constant and varies along its longitudinal direction, as shown in Eq. 4, Fig. 5, and Fig. 6.

$$A_{cross} \coloneqq A_{cross}(z) \,|\, A_{cross} : [0, L_z] \rightarrow \mathbb{R}$$

(4)

Therefore, to determine the field of resulting displacements, the Castigliano Theorem is applied. Assuming that the geometries are of the elastic-linear type, the set of forces acting on them generates a displacement field defined by the partial derivative of the total strain energy with respect to the set of forces (see Eq. 5).

$$\Delta L_z = \frac{\partial U_z}{\partial F_z}$$

(5)

Where $U_z$ [N·mm] represents the elastic or deformation potential energy generated when a linear-elastic body or element deforms a distance $\Delta L_z$ when applying a force $F_z$, in its longitudinal direction for a pure uniaxial compression service condition. Likewise, the deformation energy of a linear-elastic body or element with a variable cross-section along its longitudinal direction, in a service condition of pure uniaxial compression, can be expressed, from Eq. 2, as shown in Eq .6.

$$U_z = \int_0^{L_z} \frac{\sigma_{zz}^2}{2 \cdot E_z} \cdot dV = \int_0^{L_z} \frac{F_z^2}{2 \cdot E_z \cdot A_{cross}(z)} \cdot dz = \frac{F_z^2}{2 \cdot E_z} \cdot \int_0^{L_z} \frac{dz}{A_{cross}(z)}$$

(6)

Finally, substituting Eq. 6 in Eq. 5 and developing the expression, Eq. 7. It is possible to obtain the displacement field for a linear elastic body or element with a variable cross section along its longitudinal direction, under a pure uniaxial compression service condition.

$$\Delta L_z = \frac{\partial\left(\frac{F_z^2}{2 \cdot E_z} \cdot \int_0^{L_z} \frac{dz}{A_{cross}(z)}\right)}{\partial F_z} \rightarrow \Delta L_z = \frac{1}{2 \cdot E_z} \cdot \frac{\partial (F_z^2)}{\partial F_z} \cdot \int_0^{L_z} \frac{dz}{A_{cross}(z)} \rightarrow \Delta L_z = \frac{F_z}{E_z} \cdot \int_0^{L_z} \frac{dz}{A_{cross}(z)}$$

(7)

To validate the predictive algorithm a case study with nonlinear geometry has been defined (see Fig. 3). In this way, the resulting displacements field in the elastic area can be determined from the magnitude of the pure uniaxial compression force applied on the end of the geometry (see Fig. 3), the variation of its cross-sectional area along its longitudinal direction and the compressive Young's Modules of the filament of plastic material (see Table 1). Likewise, and with the aim that the predictive model performs the influence of the 3D printing pattern and the printing direction on the mechanical behavior of a plastic geometry under uniaxial compression loads, four different printing infill patterns (concentric, Zig – Zag, lines 0º/90º and lines 45º/45º) and 3D printing directions (Z–axis and X/Y–axis) have been analyzed.

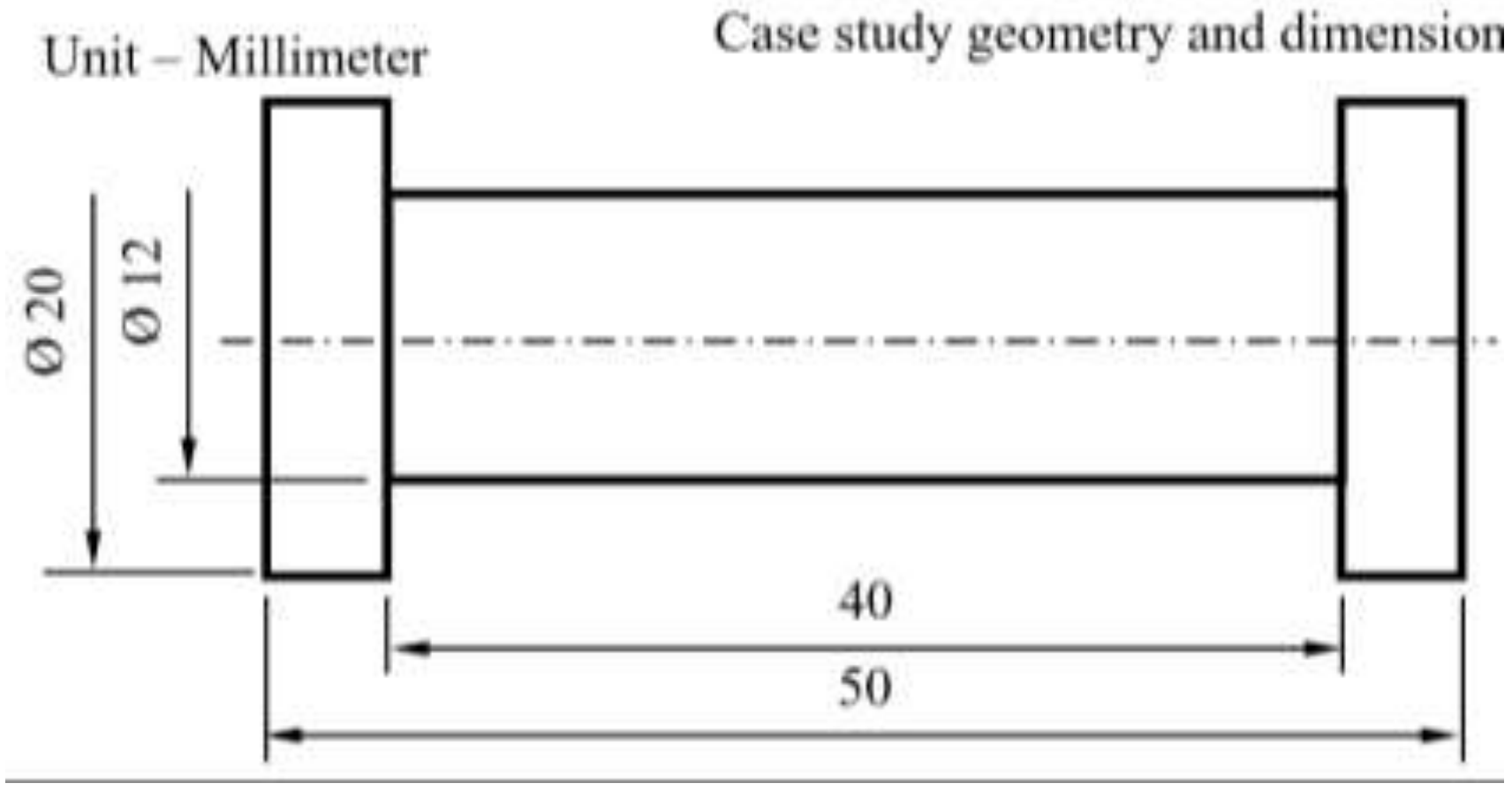

**Fig. 3** Case study geometry and dimensions

It is necessary to make a topological analysis of the part geometry to evaluate the variation of its cross-sectional area along its longitudinal direction. To do this, as shown in Fig. 4, a set of virtual parallel planes, $P_{cross} \in \mathbb{R}^3$ (see Eq. 8), whose normal direction is analogous to the longitudinal direction of the geometry and separation distance layer thickness is defined (see Table 2). The plane contours $C_{cross} \in \mathbb{R}^2$ (see Eq.8 and Fig. 4) are presented as the intersection between the set of virtual parallel planes and the geometry.

$$\forall\, P_{cross}(i) \in \mathbb{R}^3\ \exists\, C_{cross} \in \mathbb{R}^2 \mid C_{cross} = P \cap P_{cross}(i) \tag{8}$$

Being $P \in \mathbb{R}^3$ the three-dimensional model of the geometry under study (see Fig. 4). From the contours, $C_{cross} \in \mathbb{R}^2$ (see Eq. 8 and Fig. 4) it is possible to determine the cross-sectional area $A_{cross} \in \mathbb{R}^2$ for each virtual plane (see Eq. 4, Eq. 9, Fig. 5 and Fig. 6), and the magnitude $H \in \mathbb{R}$ of the layer height (see Eq. 9). Finally, as it is shown in Fig. 5, the extruder path length for Z-axis printing direction and each virtual plane is calculated taking as data the line width (see Table 1) (see Eq. 4 and Eq. 9).

$$\forall\, P_{cross}(i) \in \mathbb{R}^3\ \exists \begin{pmatrix} A_{cross}(i) \in \mathbb{R}^2 \\ H(i) \in \mathbb{R} \end{pmatrix} \Bigg| \begin{matrix} H(i) = i \cdot L_{thickness} \\ A_{cross}(i) = d_{path}(i) \cdot L_{width} \end{matrix} \tag{9}$$

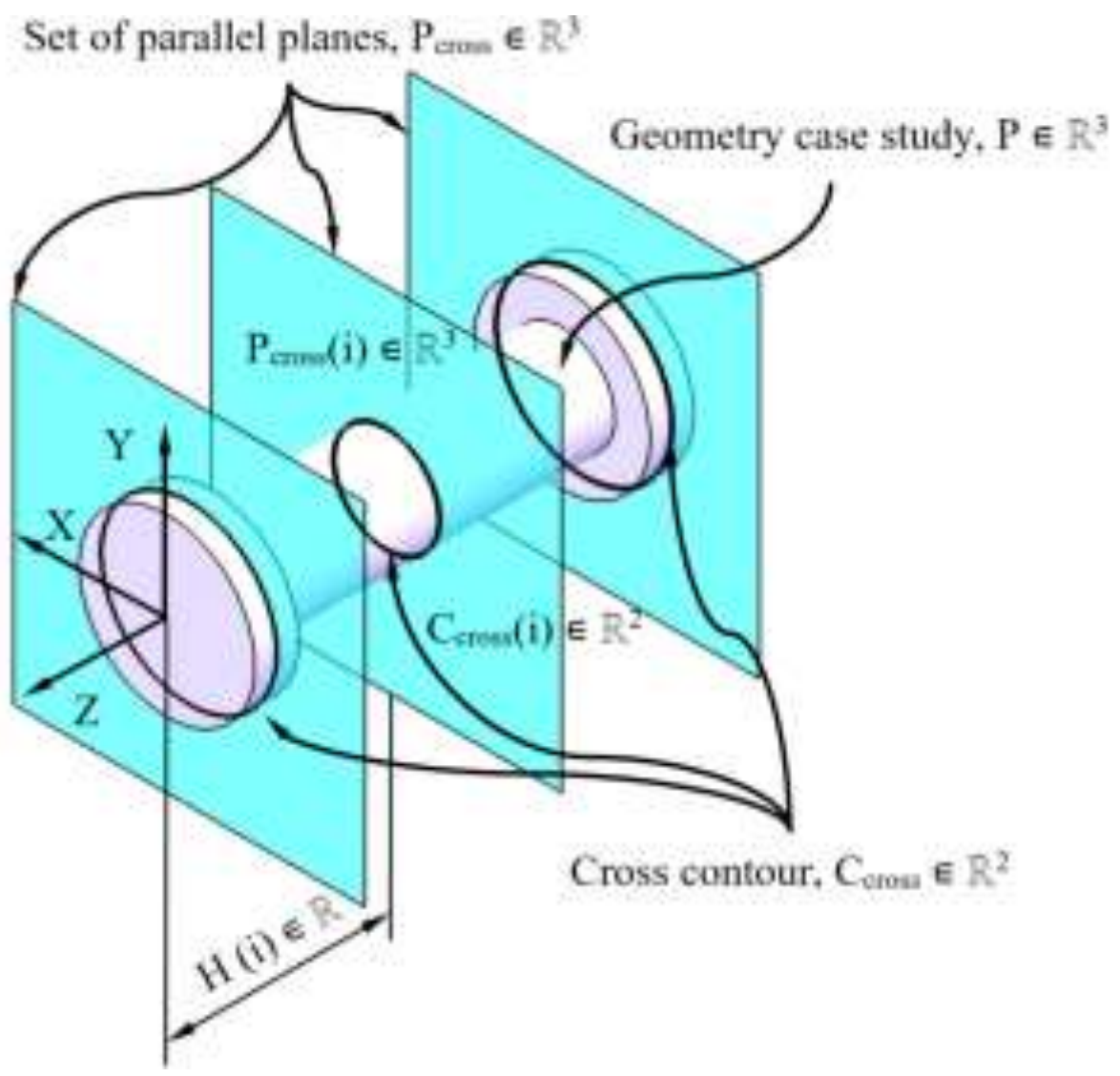

**Fig. 4** Set of parallel planes and cross contour graphic definition

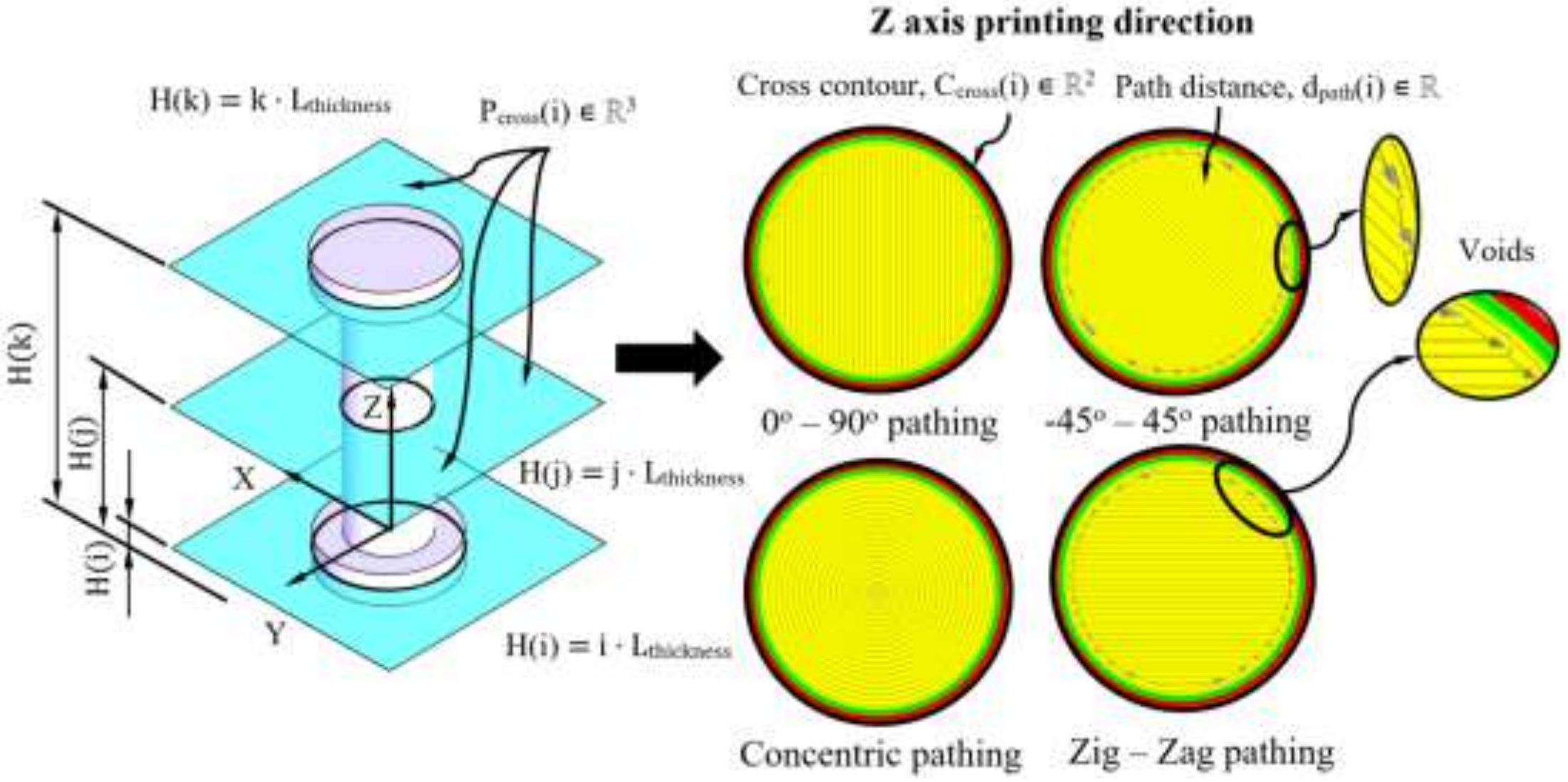

**Fig. 5** Cross area graphic definition for Z-axis printing direction

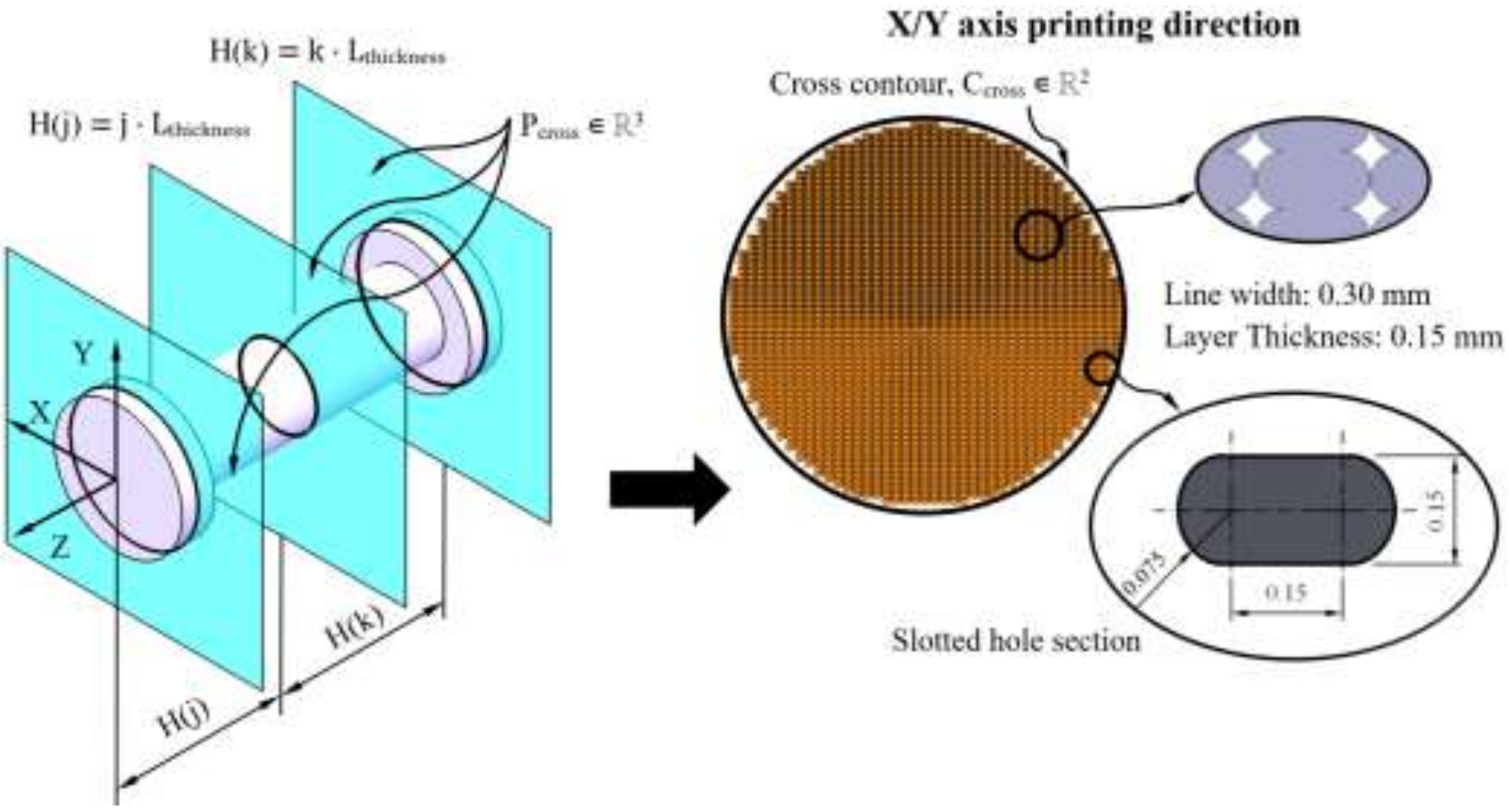

**Fig. 6** Cross area graphic definition for X/Y-axis printing direction

Where $L_{thickness}$ [mm] represents the magnitude of the geometric parameter layer thickness (see Table 2), $d_{path}$ [mm] represents the path that the extruder takes for each layer height or virtual plane defined and $L_{width}$ [mm] represents the magnitude of the geometric parameter line width (see Table 2). It should be noted that the path that the extruder makes for each layer height is established according to each of the 3D printing infill patterns analyzed and according to the geometric and technological parameters established for 3D additive manufacturing (see Table 1 and Table 2). On the other hand, for the X/Y axis printing direction, the cross-sectional area, $A_{cross} \in \mathbb{R}^2$ (see Eq. 4 and Eq. 10) is determined by evaluating the number of cross-sections of plastic material filaments, $n_{cross} \in \mathbb{R}$ (see Eq. 10), contained within each flat contour, $C_{cross} \in \mathbb{R}^2$ (see Eq. 8, Fig. 5 and Fig. 6), and the area magnitude of cross sections of plastic material filaments (see Fig. 1 and Fig. 2). All of this, for each virtual plane or defined layer height and each of the 3D printing infill patterns analyzed in this manuscript.

$$\forall\, P_{cross}(i) \in \mathbb{R}^3 \;\exists \begin{pmatrix} A_{cross}(i) \in \mathbb{R}^2 \\ H(i) \in \mathbb{R} \\ n_{cross}(i) \in \mathbb{R} \end{pmatrix} \Bigg| \begin{array}{c} H(i) = i \cdot L_{thickness} \\ A_{cross}(i) = n_{cross}(i) \cdot \left[\pi \cdot \left(\frac{L_{thickness}}{2}\right)^2 + L_{thickness}^2\right] \end{array} \quad (10)$$

As shown in Fig. 4, Fig. 5, and Fig.6 the topological analysis of the geometry under study, considers the reduction of its cross-sectional area caused by the cross-sectional holes that arise from the definition of the 3D additive manufacturing process using MEX technology. In particular, Fig.7 shows the magnitude of the cross-sectional area as a function of layer height for the geometry of the study case, $P \in \mathbb{R}^3$, and for each 3D printing infill pattern and build direction analyzed in this manuscript.

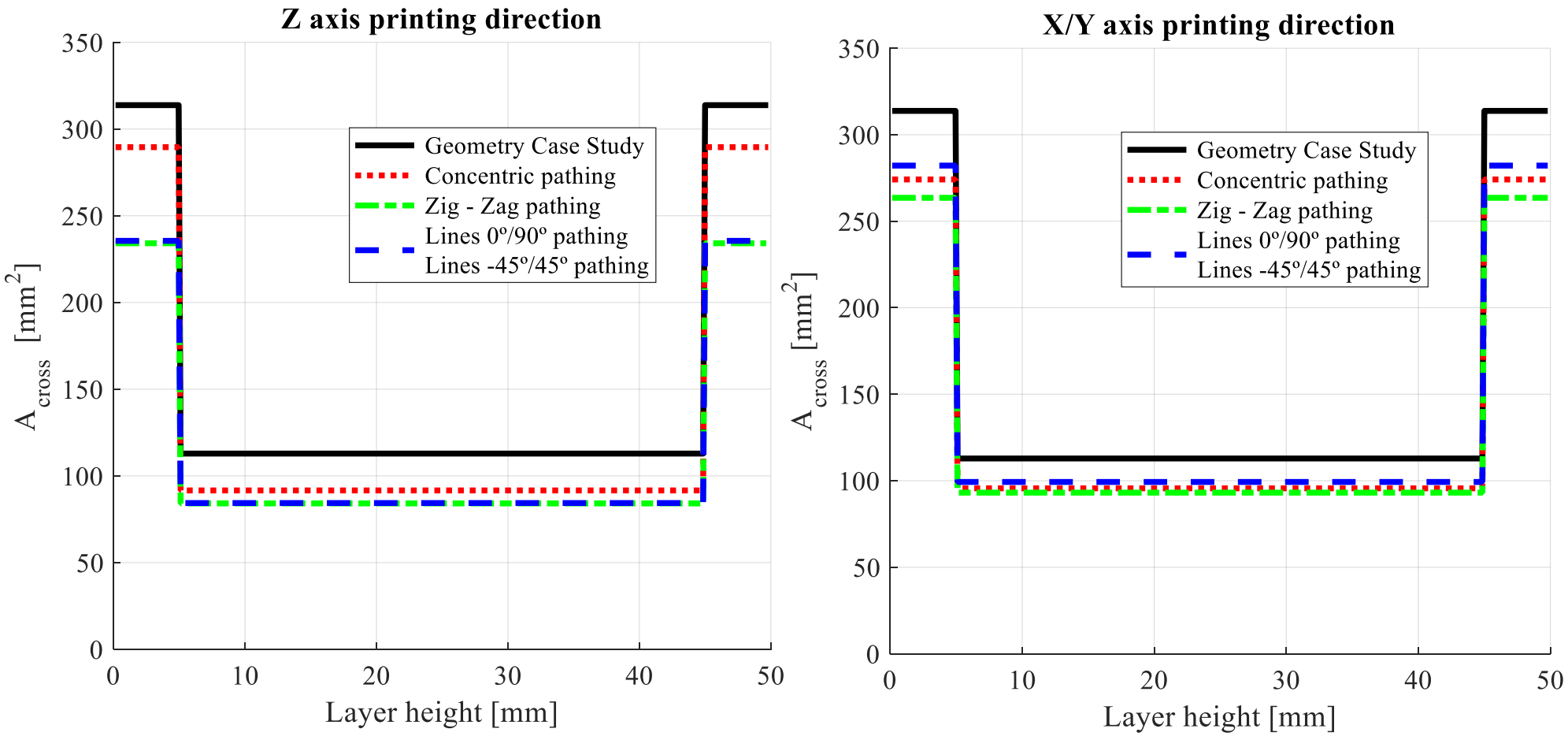

**Fig. 7** Definition of the case study geometry cross area along the layer height for the Z and X/Y axis printing direction and the used 3D printing infill patterns

In this way, once the variation of the cross-sectional area of the geometry along its longitudinal direction is known, Eq. 7 is solved (see Fig. 7) starting from a uniaxial ascending and linear pure compression force vector, determining the displacement field in the elastic-linear region of the geometry, as well as the compressive Young's modulus of the plastic material.

Annex 1 of the paper shows the flow diagram in which the geometric predictive algorithm developed to obtains the elastic behavior of the geometry under study and the Young's modulus of the plastic material under pure uniaxial compression service conditions is described. In addition, it should be noted that this geometric predictive algorithm is universal and can be applied to any geometry and plastic material used for this type of 3D additive manufacturing using MEX tecnology.

## 2.2. Experimental

For the development of the proposed methodology, a set of experimental tests have been carried out. These experimental tests are focused on characterizing the mechanical and elastic behavior of the PETG (polyethylene terephthalate glycol-modified), under pure uniaxial compression service conditions. The 3D additive manufacturing technology used is MEX and the commercial model of the 3D printer used for this manufacturing process is Ultimaker S5 (Utrecht, Netherlands) [50], see Fig. 8. This manufacturing tool allows dual extrusion of different plastic and metallic materials, with a useful printing volume of 330 x 240 x 300 mm, presenting an automatic leveling process of the hot chamber.

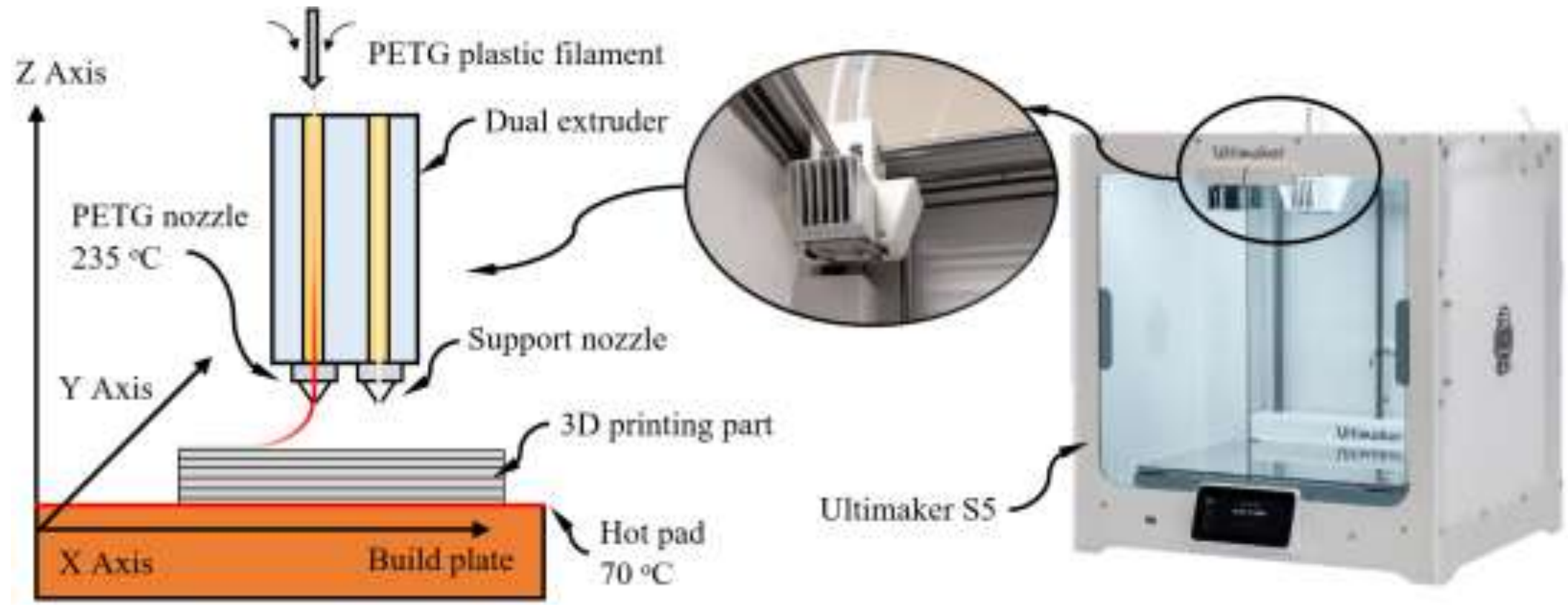

**Fig. 8** Illustration of the 3D printing manufacturing process and printing device

The configuration of the MEX manufacturing process and the selection of the main geometric and technological parameters have been carried out using commercial software Ultimaker Cura [51]. Likewise, to guarantee a high final quality of the manufactured elements, the layer height is the minimum recommended and equal to 0.15 mm, the melt temperature 235 ºC, and the hot pad temperature 70 ºC. Table 3 shows the magnitude of the rest of the geometric and technological parameters defined respecting the recommendations established by the supplier of the plastic material PETG [52].

**Table. 3** Defined geometrical and technological parameters for the 3D printing manufacturing process

| Variable | Units | Value |
|---|---|---|
| Line width | mm | 0.30 |
| Layer thickness | mm | 0.15 |
| Wall thickness | mm | 0.60 |
| Wall line count | – | 2.00 |
| Nozzle size | mm | 0.40 |
| Infill density | % | 100 |
| Printing speed | mm/s | 48 |
| Wall pattern | – | Concentric |

The original physics, thermal, and elastic filament properties [50] are shown in Table 4. As can be seen, the elastic characterization of the PETG plastic filament is orthotropic, presenting a Young's modulus of 1472 MPa along its longitudinal direction (Z–axis) and a Young's modulus of 1087 MPa along its transverse direction (X/Y–axis). This elastic orthotropic characterization of the plastic filament is considered for the theoretical and numerical modeling of the PETG plastic material using MEX technology.

**Table. 4** Properties of the PETG plastic filaments

| Properties | Units | Value |
|---|---|---|
| Material density | $g/cm^3$ | 1.27 |
| Printing temperature | ºC | 235 ± 10 |
| Hot pad temperature | ºC | 60 – 90 |
| Glass transition temperature | ºC | 81 |
| Softening temperature | ºC | 84 |
| Filament diameter | mm | 2.85 |
| Cross Young's modulus – $E_{cross}$ – X/Y–Axis | MPa | 1472 ± 270 |
| Longitudinal Young's modulus– $E_{longitudinal}$ – Z–Axis | MPa | 1087 ± 79 |

In order to perform the formal characterization of the elastic and mechanical properties of the PETG plastic material under uniaxial pure compression service conditions, the standard ASTM D695 – 15 (Standard Test Method for Compressive Properties of Rigid Plastics) [53] has been applied during the experimental procedures accomplished in the present research. Thus, the standard ASTM D695 – 15 defines the geometry of the tested specimens as cylinders which radius is 12.7 mm and height 50.8 mm, see Fig. 9.

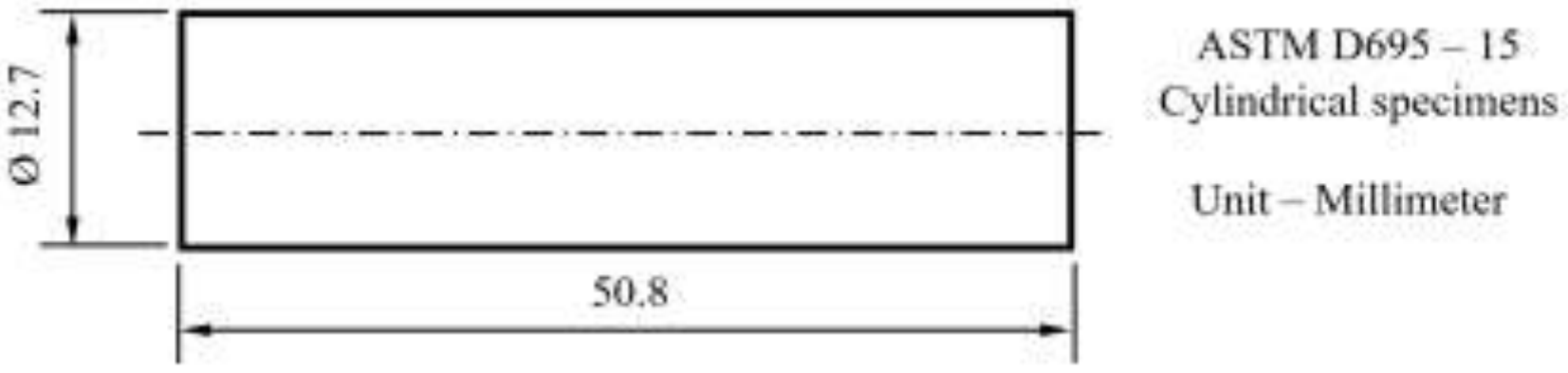

**Fig. 9** 3D printing specimen dimensions

Four different 3D printing infill patterns have been defined for the experimental tests (Concentric, Zig–Zag, Lines 0º/90º, and Lines –45º/45º), applied in the main 3D printing directions (Z–axis and X/Y–axis), see Fig. 10 and Fig. 11. In other words, for the different experimental procedures made, 8 types of test tubes have been manufactured. Furthermore, the standard ASTM D695 – 15 performs the characterization of the plastic material through the experimental test of, at least, 5 specimens. To fulfill this methodology, a total of 42 specimens have been manufactured, 6 specimens for each configuration. On the other hand, it should be noted that, given the specimens symmetry, the structural behavior of the Lines 0º/90º and Lines –45º/45º specimens, in the Z–axis direction, is analogous.

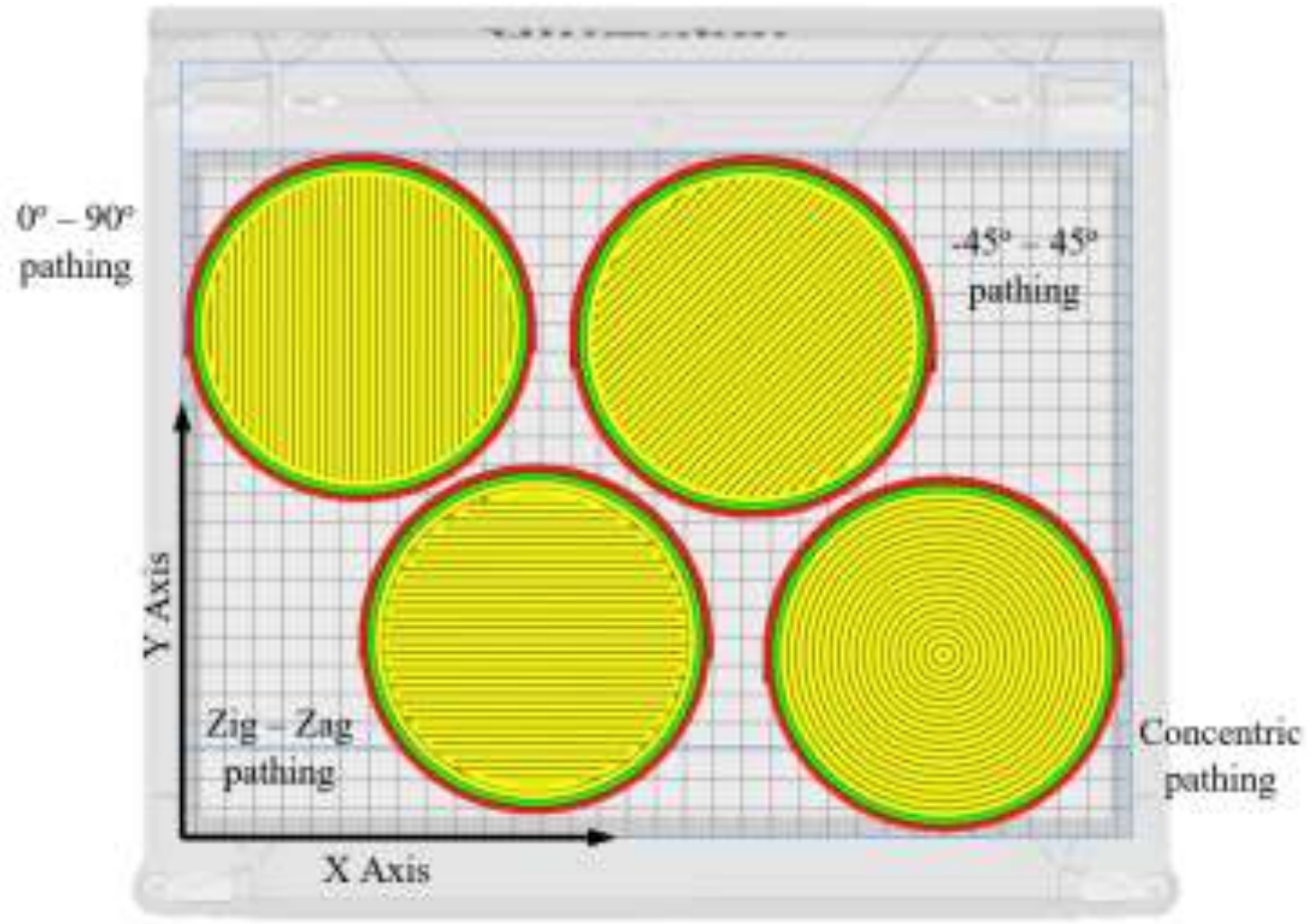

**Fig. 10** 3D printing infill patterns for Z–axis cylindrical specimens

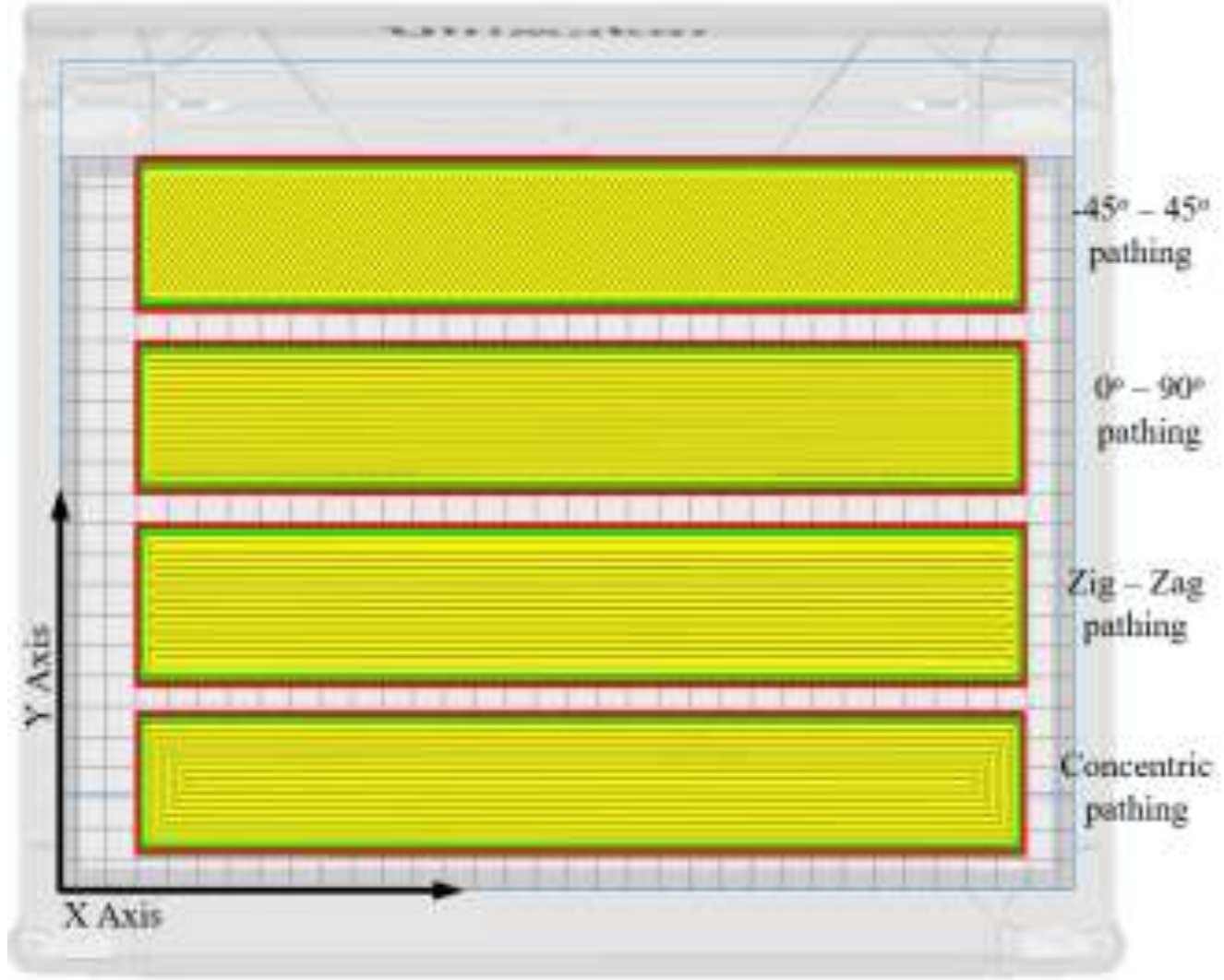

**Fig. 11** 3D printing infill patterns for X/Y–axis cylindrical specimens

In order to validate the mechanical and elastic characterization of the PETG material, as well as the methodology described in the manuscript, the case study has been tested with service conditions similar to those established for the test specimens following the standard ASTM D695 – 15. As shown in Fig. 12, the case study presents a topology with nonlinear geometry, which also allows evaluating the influence of this variable on the elastic and structural behavior. In order to establish a correlation between the results obtained after the experimental tests of the specimens and the case study, the geometric and technological parameters used in their manufacture are analogous. To evaluate the influence of the 3D printing infill pattern parameter (Concentric, Zig–Zag, Lines 0º/90º and Lines –45º/45º) and the main manufacturing direction (Z–axis and X/Y–axis), eight different models for this case study have been manufactured. To improve the final quality of the manufactured parts the option to include supports, when necessary, has been implemented, see Fig. 5 and Fig. 6.

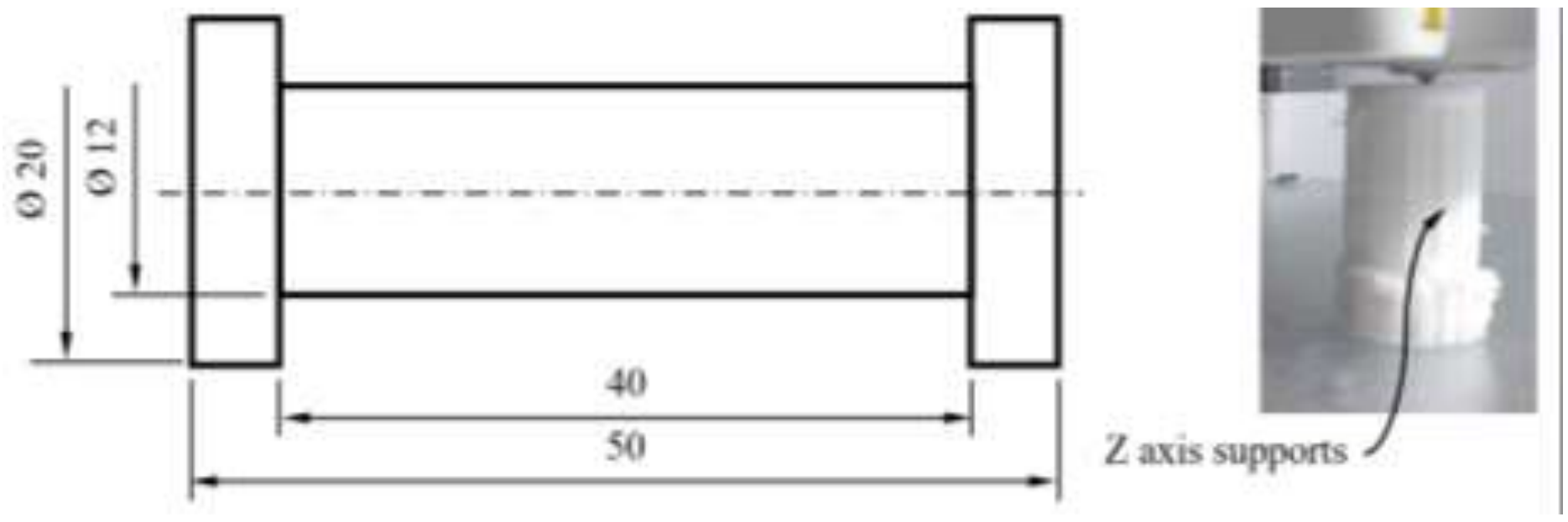

**Fig. 12** Case study manufacturing

The definition of the compression test to obtain the parameters, compressive strength and compressive Young's Modulus for the printed plastic material PETG is established in the standard ASTM D695 – 15 (Standard Test Method for Compressive Properties of Rigid Plastics). The test device to carry out the uniaxial pure compression tests is MTS – 810 (Minnesota, United States) [54] see Fig 13. According to the requirements of the standard, the device has two supports, one mobile and the other fixed, anchoring the flat jaws that generate the unidirectional compression movement at a constant speed. The compression speed for all tests is 1.3 ± 0.3 mm/min. This test device has a compressometer and a load indicator to record the distance traveled between both clamps and the compression force applied, for any time during the experimental tests.

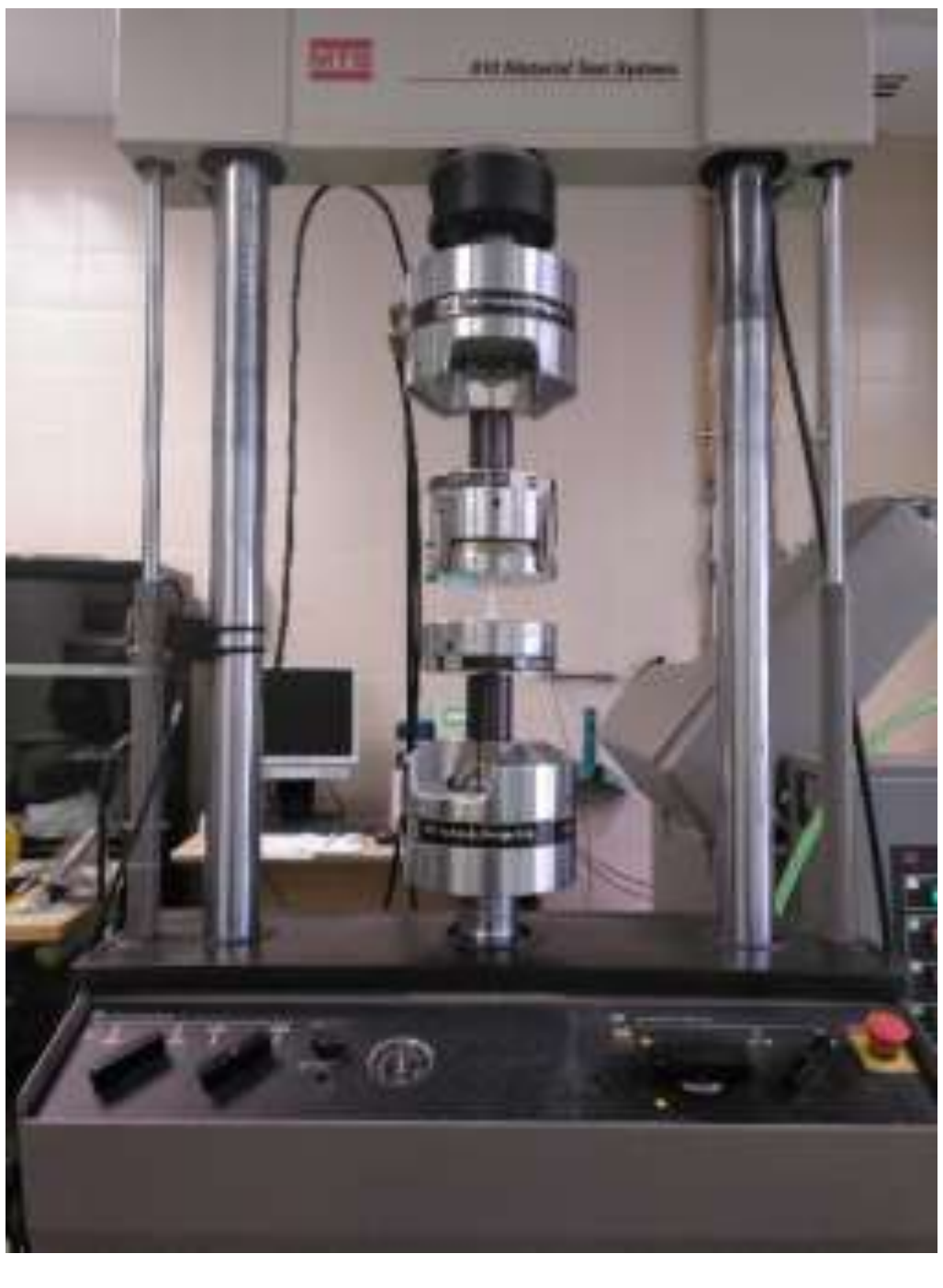

**Fig. 13** Schematic of the uniaxial compression test

2.3. Numerical

In order to compare and validate the results of the geometric predictive algorithm and the different experimental tests carried out, a set of numerical analyzes, for the geometry under study, through FEM simulations using the numerical CAE Ansys Mechanical [55] have been proposed. Fig. 14 shows the load scenario and the boundary conditions from which the structural and mechanical behavior of the geometry is numerically analysed. On the one hand, for the load scenario, a uniaxial compression force is established with a direction aligned with the longitudinal axis of the geometry, applied on its upper horizontal surface and with a magnitude of 600 N (see Fig. 14). On the other hand, for the boundary conditions, a fixed support or embedment located on the lower horizontal surface is defined, that is, the support base of the geometry (see Fig. 14).

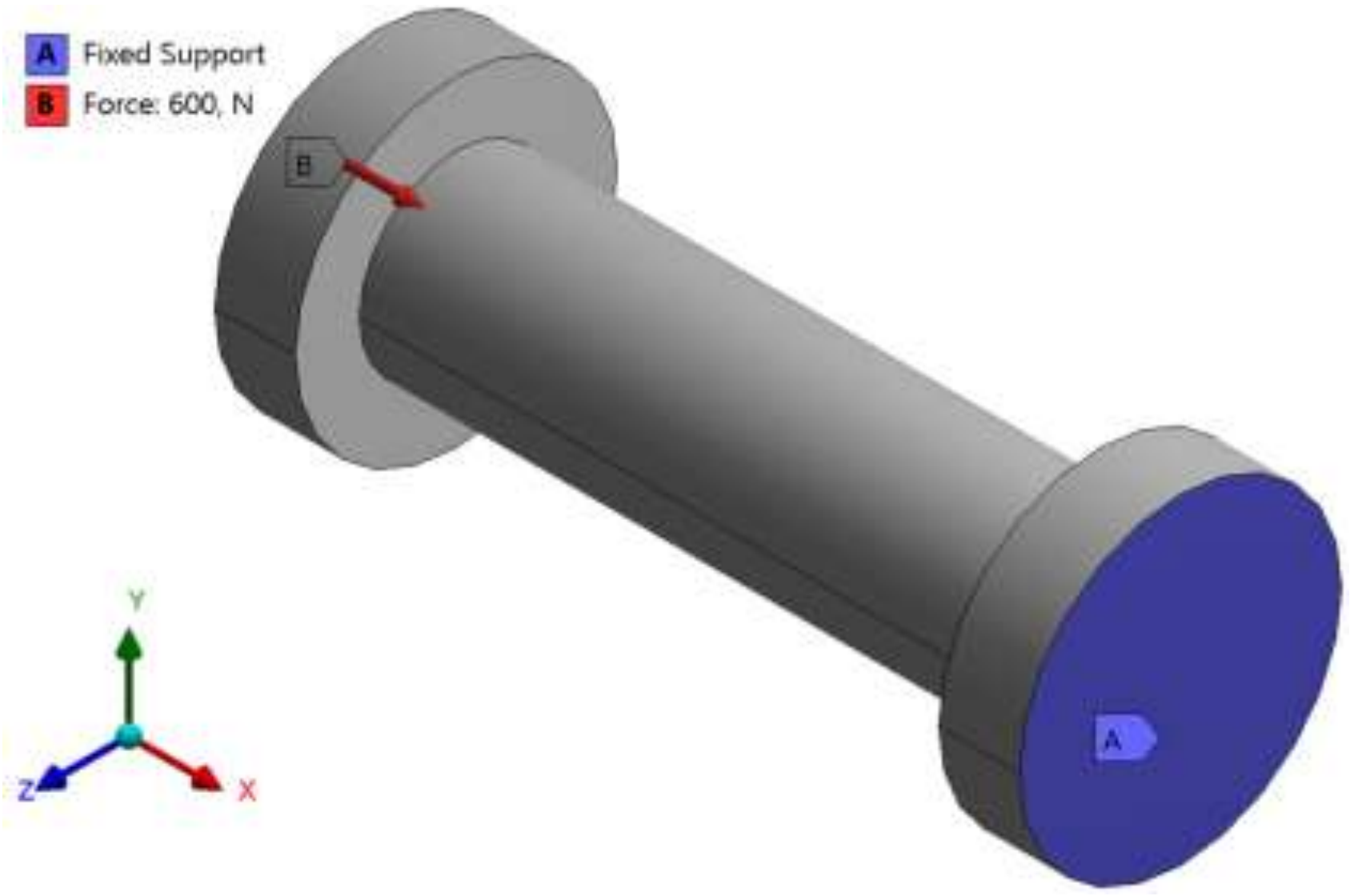

**Fig. 14** Boundary conditions and loads scenario for the numerical simulation of the geometry case study

The definition of the solver for the numerical model has been elastic–linear. In total, seven numerical simulations have been carried out to evaluate the mechanical and structural behavior of the geometry for each of the 3D printing infill patterns, Concentric, Zig–Zag, Lines 0º/90º and Lines –45º/45º, and for each 3D printing direction, Z and X/Y–axis (see Fig. 10 and Fig. 11). The main objective of the numerical simulations is to establish a comparison between the numerical results, the results of the predictive model and the results obtained from the experimental tests. In this way, it can be verified if the elastic and mechanical characterization carried out for the PETG plastic material can be applied in numerical models for the mechanical analysis of 3D printing geometries manufactured by MEX additive tecnology. For all the numerical simulations carried out, the PETG plastic material has been defined as an isotropic, elastic, and linear material, with a Poisson's ratio equal to 0.38. For the compressive Young's modulus, the values of this elastic parameter obtained from the experimental tests carried out have been used . Finally, for the meshing process, tetrahedral elements of the SOLID 92 type have been defined. This type of finite element is second order and is composed of 10 nodes (see Fig. 15). The accuracy used in the meshing process is 0.5 mm. Table 5 and Fig. 15 show the number of nodes and number of elements defined for the mesh, as well as the average quality of the tetrahedral elements.

**Table. 5** Mesh statistics defined for the numerical simulations

| Number of elements | Number of nodes | Element quality (average) |
|---|---|---|
| 520,540 | 721,417 | 0.849 |

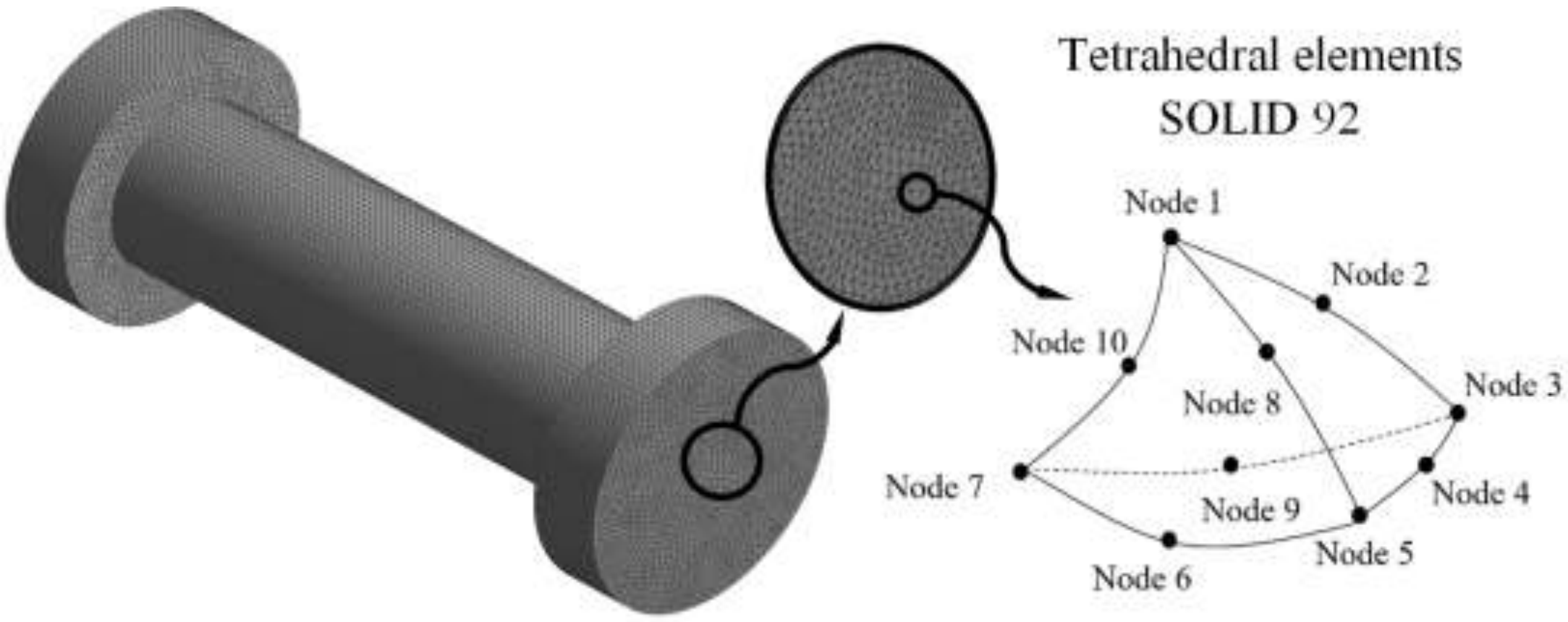

**Fig. 15** Mesh definition for the numerical simulation of the geometry case study

## 3. Results and discussions

### 3.1. Compressive Young's modulus characterization for the PETG plastic material

Table 6, and Table 7 show the experimental results obtained from the tests for each type of cylindrical specimen analyzed. In addition, for each 3D printing infill pattern, Concentric, Zig–Zag, Lines 0º/90º and Lines –45º/45º, and for each 3D printing direction, Z and X/Y–axis, compressive stress – nominal strain curves have been generated and included in the Appendix A of the present manuscript, see Fig. 23 and Fig. 24. The results of uniaxial compressive force and the field of nominal displacements have been obtained by means of the measuring instruments of the testing machine (see Fig. 13) [52]. All this is, according to the methodology, described in the international standard ASTM D695 – 15 [51]. In this way, from the uniaxial pure compressive stress field and the resulting nominal strain field (see Fig. 23 and Fig. 24) it is possible to determine the elastic and mechanical parameters of compressive yield strength, ultimate stress, and compressive Young's modulus, for the plastic material PETG. Table 4 and Table 5 show the magnitude of these parameters for each configuration defined in the 3D additive manufacturing process using MEX technology, including statistical values of std deviation and variance.

**Table. 6** Elastic and mechanical compression properties for the Z–axis printed specimens

| Mechanical properties | Concentric pathing | Zig – Zag pathing | Lines 0º/90º & 45º/–45º pathing |
|---|---|---|---|
| **Young's modulus** [MPa] | 1,018.0 | 1,141.4 | 1,030.0 |
| Std deviation | 16.2 | 25.1 | 18.7 |
| Variance | 315.2 | 632.2 | 468.2 |
| **Yield strength** [MPa] | 20.67 | 26.06 | 11.64 |
| Std deviation | 0.94 | 1.36 | 1.67 |
| Variance | 1.07 | 1.86 | 2.80 |
| **Ultimate stress** [MPa] | 33.53 | 39.02 | 27.15 |
| Std deviation | 0.76 | 1.26 | 0.76 |
| Variance | 0.70 | 1.57 | 0.58 |

**Table. 7** Elastic and mechanical compression properties for the X/Y–axis printed specimens

| Mechanical properties | Concentric pathing | Zig – Zag pathing | Lines 0º/90º pathing | Lines 45º/–45º pathing |
|---|---|---|---|---|
| **Young's modulus** [MPa] | 1,333.2 | 1,273.9 | 1,326.5 | 1,252.7 |
| Std deviation | 21.5 | 13.7 | 16.2 | 14.7 |
| Variance | 577.5 | 234.9 | 380.2 | 231.6 |
| **Yield strength** [MPa] | 29.16 | 22.41 | 27.28 | 33.83 |
| Std deviation | 2.79 | 1.18 | 1.00 | 1.17 |
| Variance | 9.71 | 1.73 | 0.95 | 1.72 |
| **Ultimate stress** [MPa] | 43.24 | 34.51 | 36.24 | 40.05 |
| Std deviation | 2.32 | 0.72 | 0.76 | 0.44 |
| Variance | 6.71 | 0.64 | 0.83 | 0.24 |

On the other hand, Table 8 and Table 9 show the results obtained from the predictive algorithm described in this manuscript. The compressive Young's modulus of the PETG plastic material is determined by the compression stress – nominal strain ratio derived, in turn, from the vector of uniaxial compression forces and the displacement field (see Eq. 7) that it generates on the geometry of the cylindrical test specimens. In this way, the compressive Young's modulus and the elastic–linear region of a plastic material that will be used in the 3D additive manufacturing of a part can be determined. The predictive methodology presented does not require the performance of experimental tests for the mechanical and elastic characterization of the plastic material used in the 3D active manufacturing process, reducing manufacturing costs and subsequent experimental testing for each configuration proposed in the 3D additive manufacturing of a piece.

**Table. 8** Compressive Young’s modulus obtained through the predictive algorithm for the Z–axis printing direction

| 3D infill patterns | Compressive Young’s modulus [MPa] | | Relative error [%] |
|---|---|---|---|
| | Predictive algorithm | Experimental tests | |
| Concentric | 1,068.9 | 1,018.0 | 4.99 |
| Zig – Zag | 1,080.7 | 1,141.4 | 5.32 |
| Lines 0º/90º & 45º/–45º | 1,090.5 | 1,030.0 | 5.87 |

**Table. 9** Compressive Young’s modulus obtained through the predictive algorithm for the X/Y–axis printing direction

| 3D infill patterns | Compressive Young’s modulus [MPa] | | Relative error [%] |
|---|---|---|---|
| | Predictive algorithm | Experimental tests | |
| Concentric | 1,298.3 | 1,333.2 | 4.82 |
| Zig – Zag | 1,212.5 | 1,273.9 | 2.62 |
| Lines 0º/90º | 1,291.1 | 1,326.5 | 2.67 |
| Lines 45º/–45º | 1,203.3 | 1,252.7 | 3.94 |

Moreover, Table 8 and Table 9 compare the predictive results with the experimental ones (see Table 4 and Table 5). As can be seen, the relative error associated with the results obtained through the predictive model ranges from 2.62% ~ 5.87%, in comparison to the results obtained from the experimental tests for each configuration of the 3D additive manufacturing process. In this way, from the resulting relative error, the predictive algorithm can be validated, as well as the predictive determination of the compressive Young's modulus based on the geometric and technological parameters defined during the configuration of the 3D additive manufacturing process. It should be noted that the predictive algorithm presented in this manuscript is universal and can be applied to any plastic material and any part manufactured using the MEX additive manufacturing process.

Based on the experimental and predictive results it is verified that, in general and for all the 3D printing patterns analyzed, the manufacturing direction that presents the best elastic and structural behavior is the X/Y direction. As shown in Table 6, Table 7, Table 8, and Table 9, in the X/Y printing direction the magnitude of the parameters compressive Young's modulus, yield strength, and ultimate strength is higher than that obtained for the printing direction Z. The main reason that justifies the improvement of the elastic and structural behavior of the PETG plastic material in the X/Y–axis printing direction is that, for this direction, most part of the material filament is longitudinally aligned with the direction of load scenario application. While the plastic filament exhibits a higher Young's modulus in its transverse direction, as indicated in Table 6, aligning the compression force with the primary filament direction in the specimens, coupled with the reduction in the number of voids (refer to Fig. 7), results in an enhanced structural and elastic performance of the PETG plastic material along the X/Y-axis printing direction.

Based on both experimental findings and predictive outcomes, it is established that the 3D concentric infill pattern emerges as the configuration with superior elastic and structural characteristics. In conjunction with the two examined printing directions, this particular print pattern demonstrates a higher magnitude of compressive Young's modulus, yield strength, and ultimate strength parameters in comparison to the other print patterns. Ultimately, it is important to highlight that, considering the geometric condition of revolution inherent in the analyzed cylindrical specimens, the experimental outcomes for the primary manufacturing direction Z exhibit similarity in the Lines –45º/45º and Lines 0º/90º printing patterns. In this instance, the structural performance of both 3D printing configurations is essentially identical.

Moreover, the methodology presented in this manuscript has been applied and validated through a parallel case study, which was previously conducted by the same authors [56]. Table 10 shows the experimental and predictive results of the mechanical characterization of the PLA plastic material subjected to a tensional state of pure uniaxial compression. In this case study, rectangular specimens are used to determine the Young's Modulus, by the ISO-604 standard [57]. As can be seen from the analysis, the outcomes of both the experimental and predictive results, derived from the proposed algorithm, exhibit equivalence, with a confidence level or relative error standing at only 0.4%.

Ultimately, according to [58-64] the mechanical and elastic properties of PETG and PLA polymer materials differ in magnitude, although their mechanical and elastic behavior is analogous. Specifically, as in [58-64], the structural behavior of the polymers analyzed, under a tensional state of pure uniaxial compression, begins with an elastic deformation of the component until reaching the yield stress, followed by strain softening, strain hardening, and eventual fracture or structural collapse, as shown in Figure 24 and 25. The analytical results obtained from the proposed methodology allow modeling the elastic behavior of polymer materials that, applied to the geometries of the plastic specimens, determine the elastic modulus according to the ASTM D695 – 15 standard and which, in turn, can be used for mechanical simulation processes, as well as previous studies [58-64].

On the other hand, other studies [65-67] have established predictive models to evaluate the mechanical and elastic behavior of polymer materials. The proposed methodology stands out for its analytical and parameterized approach. Once validated, the model does not require additional experimental verification. This feature not only saves time and material, but also contributes to energy efficiency, highlighting its resource efficiency. Eliminating the need for additional experimental validations significantly reduces the material and energy costs associated with the experimental analysis approach. Furthermore, its parameterization allows to efficiently evaluating the elasticity of the component in the face of possible changes in the geometric and technological parameters of the MEX additive manufacturing process, without the need to carry out additional experimental tests.

**Table 10.-** Application of the proposed methodology for the PLA plastic material and test specimens with rectangular geometry [54].

| Method Uniaxial Compression Test | Young's Modulus [MPa] |
|---|---|
| Experimental | 2528.5 |
| Predictive | 2538.7 |
| Relative Error [%] | 0.403 |

3.2. Geometry case study compressive analysis

After conducting elastic characterization of the PETG plastic material, utilizing data from experimental tests on cylindrical specimens and results from the presented predictive algorithm, the mechanical behavior of the studied geometry is thoroughly examined. Table 11, and Table 12 show, for the geometry under study, the magnitude of the resulting nominal displacement obtained for a uniaxial compression force of 600 N (see Fig. 14), applied on the end of the part. Moveover, Table 11 and Table 112 show the magnitude of the nominal displacement obtained from the predictive algorithm presented and the numerical simulations performed, for the same service condition. Moreover, uniaxial compression force–nominal displacement curves have been generated for each 3D printing infill pattern and direction. These curves, specific to the geometry case study, are detailed in Appendix A of the present manuscript, as illustrated in Fig. 25 and Fig. 26. As can be verified, the results obtained by the predictive algorithm are close to the experimental results with a relative error of 0.2278% ~ 4.3684%. Therefore, the application of the predictive algorithm proposed can be validated, since the relative error committed in this approach does not exceed 5%, compared to the experimental solution obtained. However, for the results obtained through the numerical simulations, the relative error made in the approximation is greater and varies between 1.137% ~13.555%. The main reason why the numerical results do not approximate with such precision, in some 3D additive manufacturing configurations, with the experimental results is the evaluation of the volume of holes that the geometry under study presents. That is to say, to define the elastic properties of the PETG plastic material, in the numerical simulations, the experimental results of the cylindrical specimens have been used, however, the geometry under study does not present the same volume distribution of holes as these. Therefore, without evaluating the actual volume of holes that the geometry presents, and by importing the solid CAD model, the relative error obtained, in some 3D additive manufacturing configurations, is greater than 5% and is far from the experimental results obtained.

**Table. 11** Nominal displacement obtained for the geometry case study printed in the Z–axis direction

| 3D infill patterns | Nominal displacement [mm] | | | Relative error [%] | |
|---|---|---|---|---|---|
| | Experimental result | Predictive result | Numerical result | Predictive model | Numerical simulation |
| Concentric | 0.2634 | 0.2640 | 0.2732 | 0.228 | 3.717 |
| Zig – Zag | 0.2747 | 0.2867 | 0.2437 | 4.368 | 11.303 |
| Lines 0º/90º & 45º/–45º | 0.2934 | 0.2834 | 0.2700 | 3.408 | 7.972 |

**Table. 12** Nominal displacement obtained for the geometry case study printed in the X/Y–axis direction

| 3D infill patterns | Nominal displacement [mm] | | | Relative error [%] | |
|---|---|---|---|---|---|
| | Experimental result | Predictive result | Numerical result | Predictive model | Numerical simulation |
| Concentric | 0.2110 | 0.2098 | 0.2086 | 0.569 | 1.137 |
| Zig – Zag | 0.2270 | 0.2309 | 0.2183 | 1.718 | 3.8282 |
| Lines 0º/90º | 0.1966 | 0.2033 | 0.2097 | 3.408 | 6.638 |
| Lines 45º/–45º | 0.1955 | 0.1922 | 0.2220 | 1.688 | 13.555 |

Finally, from the results presented in Table 11 and Table 12, it is determined that the nominal displacement magnitudes obtained for the X/Y–axis printing direction are smaller than those obtained for the Z–axis printing direction. In this way, it is verified that the structural and mechanical behavior of the geometry under study improves for the X/Y–axis printing direction, since the elastic and mechanical properties obtained for the PETG plastic material are better, as it is shown in Table 9 and Fig. 24.

## 5. Fractology

To complete the experimental analysis of the PETG plastic material, its fracture typology and structural failure by using a high-resolution scanning electron microscope (FESEM) is evaluated. As shown in Fig. 16, the fractology analysis has been carried out for each configuration defined in the 3D additive manufacturing process of the cylindrical specimens tested. The commercial model of the microscope used is Carl Zeiss: Merlin. This measurement device has EDX and WDX Oxford analytical capabilities, a hot-tip laser gun for electron field emission, high-resolution secondary electron detectors located in the measurement chambers (SE – Everhart – Tornley and SE in lens), high-resolution back-driven electron detectors located in the measurement chambers (4-quadrant solid-state AsB integrated into the lens of the GEMINI II column and EsB in lens) and cathodoluminescence (CL) detectors. The maximum image resolutions offered by this device are 0.8 nm at 15 kV, 1.4 nm at 1 kV, and 2.4 nm at 0.2 kV. Likewise, the magnitude of its potential acceleration range is between 0.02 V and 30 kV and the electron beam current between 10 pA and 300 nA.

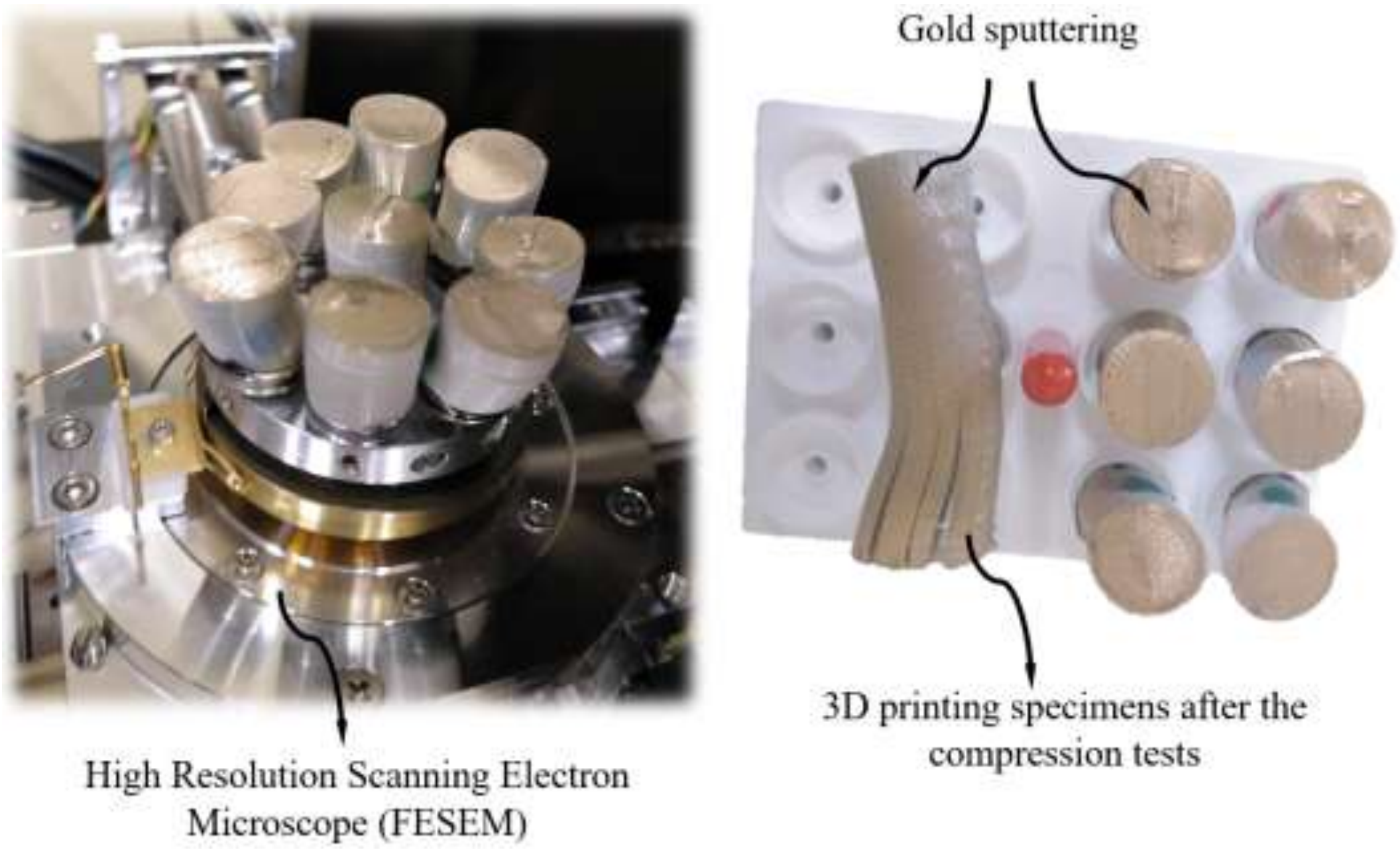

**Fig. 16** Set–up for the fractology analysis

As it is shown in Fig. 16 the specimens analyzed, by microscopy, have been covered with a thin layer of gold to give them conductive properties and improve picture conditions. In this way, it is possible to generate a beam of backscattered (e1) and secondary (e2) electrons.

To evaluate the fracture mechanics of the PETG plastic material, five test samples have been selected for cylindrical specimens manufactured in the Z and X/Y–axis printing directions with concentric, Zig–Zag and lines 3D infill patterns. As shown in Fig. 17, Fig. 18 and Fig. 19, the fracture presented by the cylindrical specimens manufactured in the Z axis printing direction is brittle and is generated mainly by delamination between adjacent layers of filament. In particular, for this 3D additive manufacturing configuration, the structural collapse of PETG plastic material begins when it reaches the compressive Yield strength (see Table 4, Table 5 and Fig. 23 and Fig. 24). From the magnitude of this parameter, the cylindrical specimens begin to harden through the plastic deformation of their central area. As shown in Fig. 22, the plastic deformation causes an eccentricity in the application of the uniaxial compression force, generating shear and bending stresses on the central plastic filament layers and causing the process of delamination and final breakage of the central layers. In addition, during the delamination process some plastic filaments suffer brittle fractures (see Fig. 17, Fig. 18 and Fig. 19) caused by the shear and bending stresses derived from the permanent plastic deformation suffered by the cylindrical specimens (see Fig. 22).

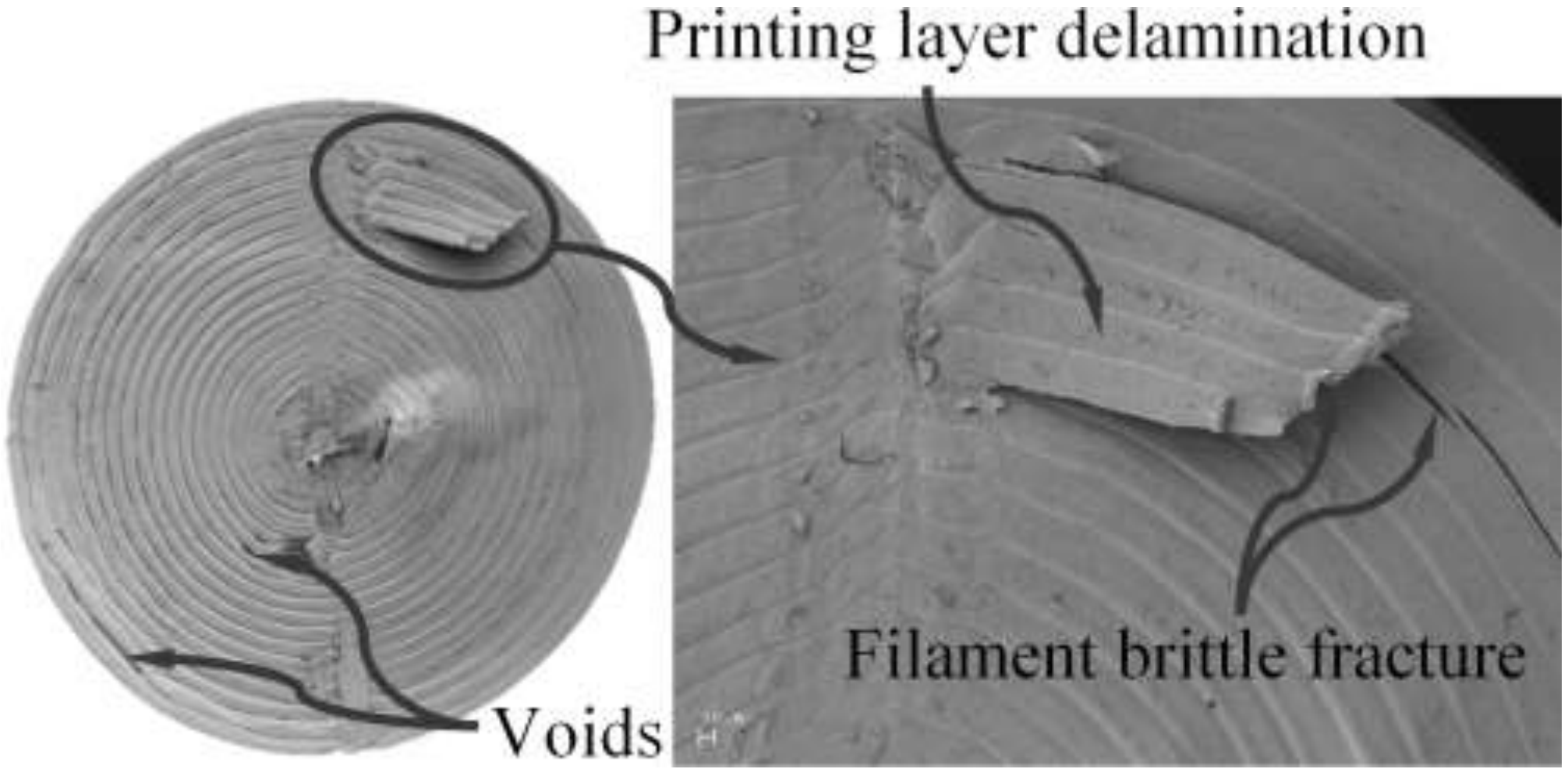

**Fig. 17** FESEM fracture images for Z axis printing direction and concentric infill pathing specimens

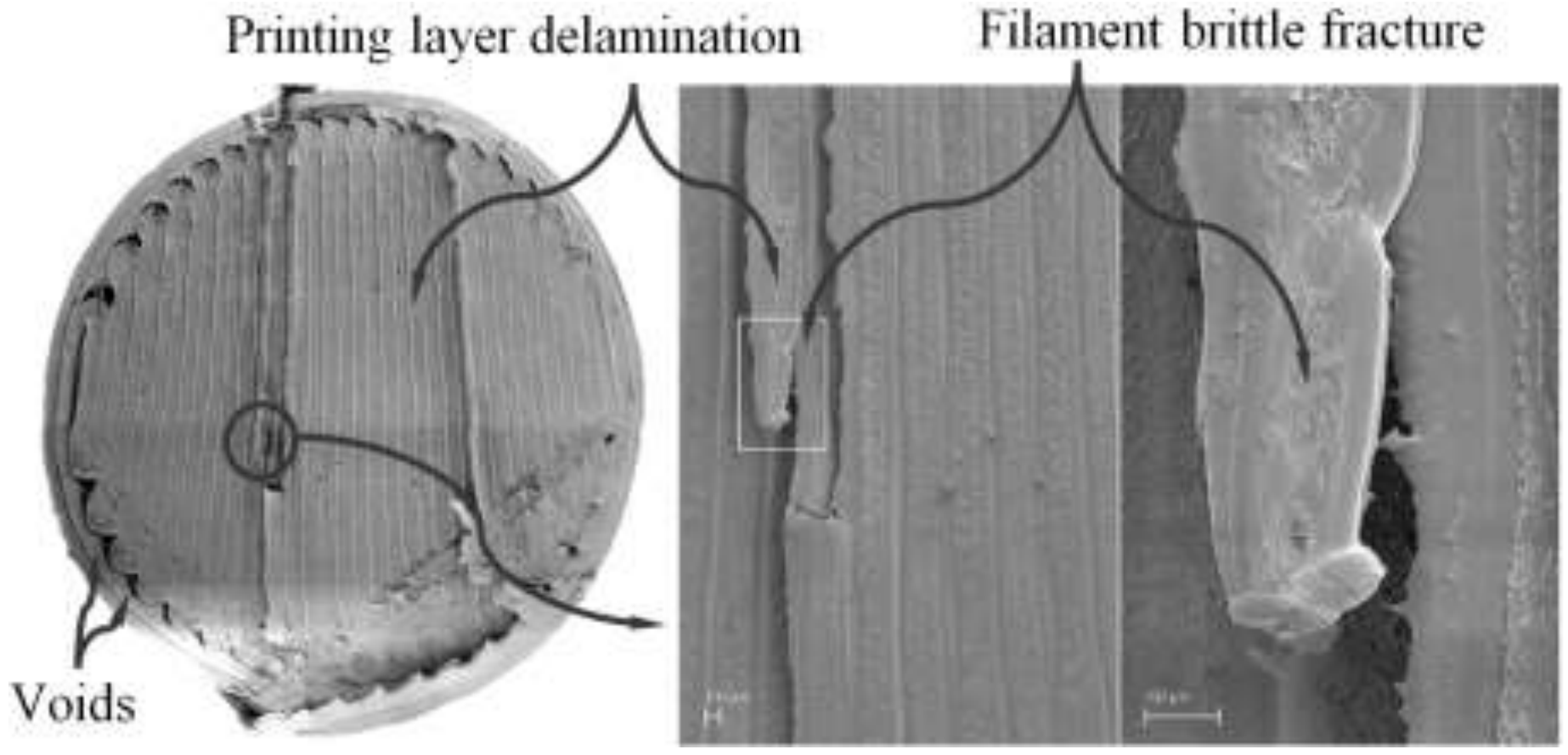

**Fig. 18** FESEM fracture images for Z axis printing direction and Zig – Zag infill pathing specimens

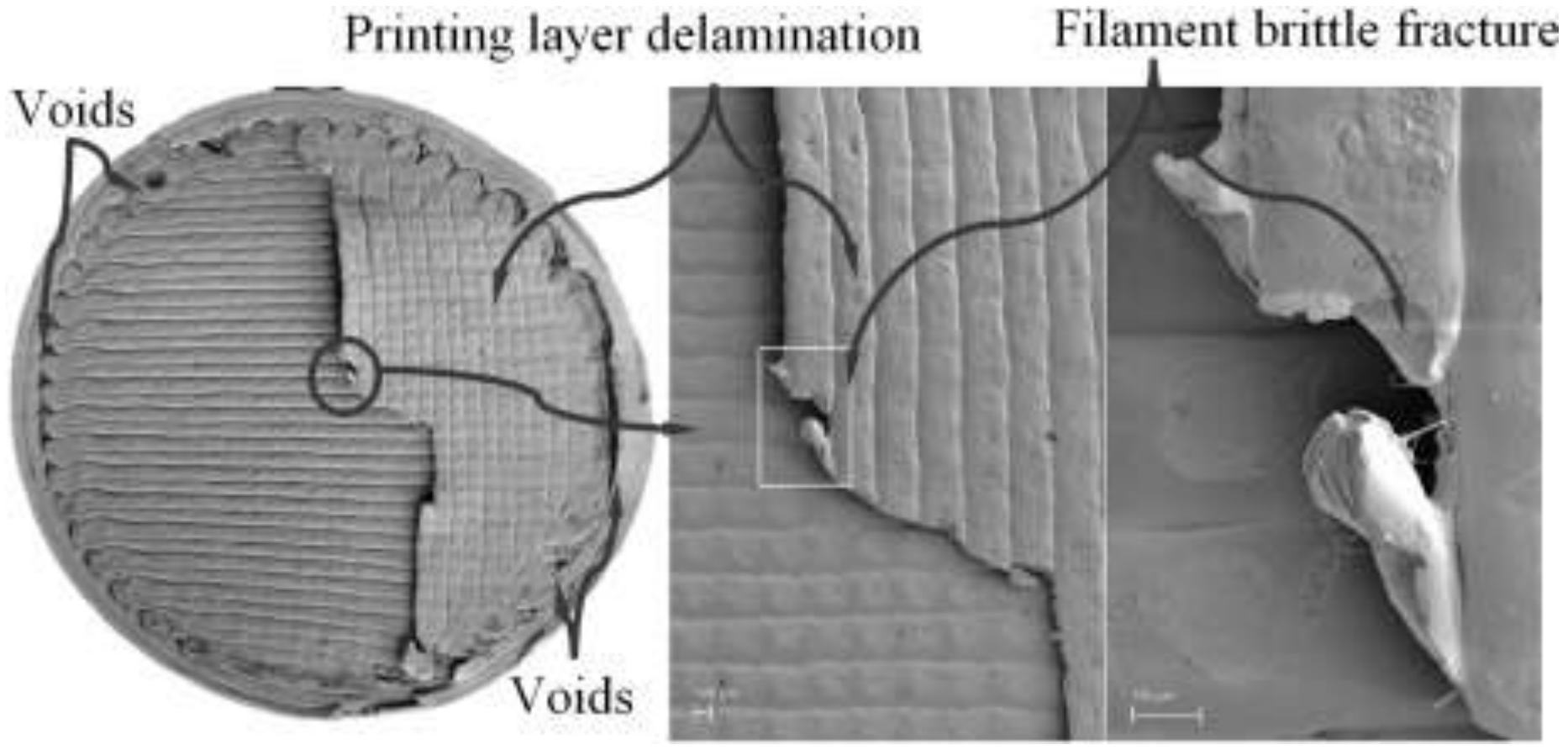

**Fig. 19** FESEM fracture images for Z axis printing direction and Lines 0º/90 & 45º/–45º infill pathing specimens

Finally, Fig. 20 and Fig.21 shows the resulting fractology for the cylindrical specimens fabricated in the X/Y–axis printing direction. As can be seen, the type of fracture generated for this set of cylindrical specimens is brittle and is mainly caused by delamination between adjacent layers of PETG plastic material. However, unlike cylindrical specimens fabricated in the Z–axis printing direction, cylindrical specimens fabricated in the X/Y–axis printing direction do not show hardening from their own plastic deformation. In other words, when they reach the compressive yield strength, crack growth is generated between contiguous layers of plastic filament that rapidly weaken the structural integrity of the cylindrical specimens. Thus, it quickly causes structural failure, without allowing a possible hardening of the material through plastic deformation. It should be noted that this type of fracture is repeated and is analogous to each 3D infill pattern defined in the 3D additive manufacturing of the cylindrical test specimens.

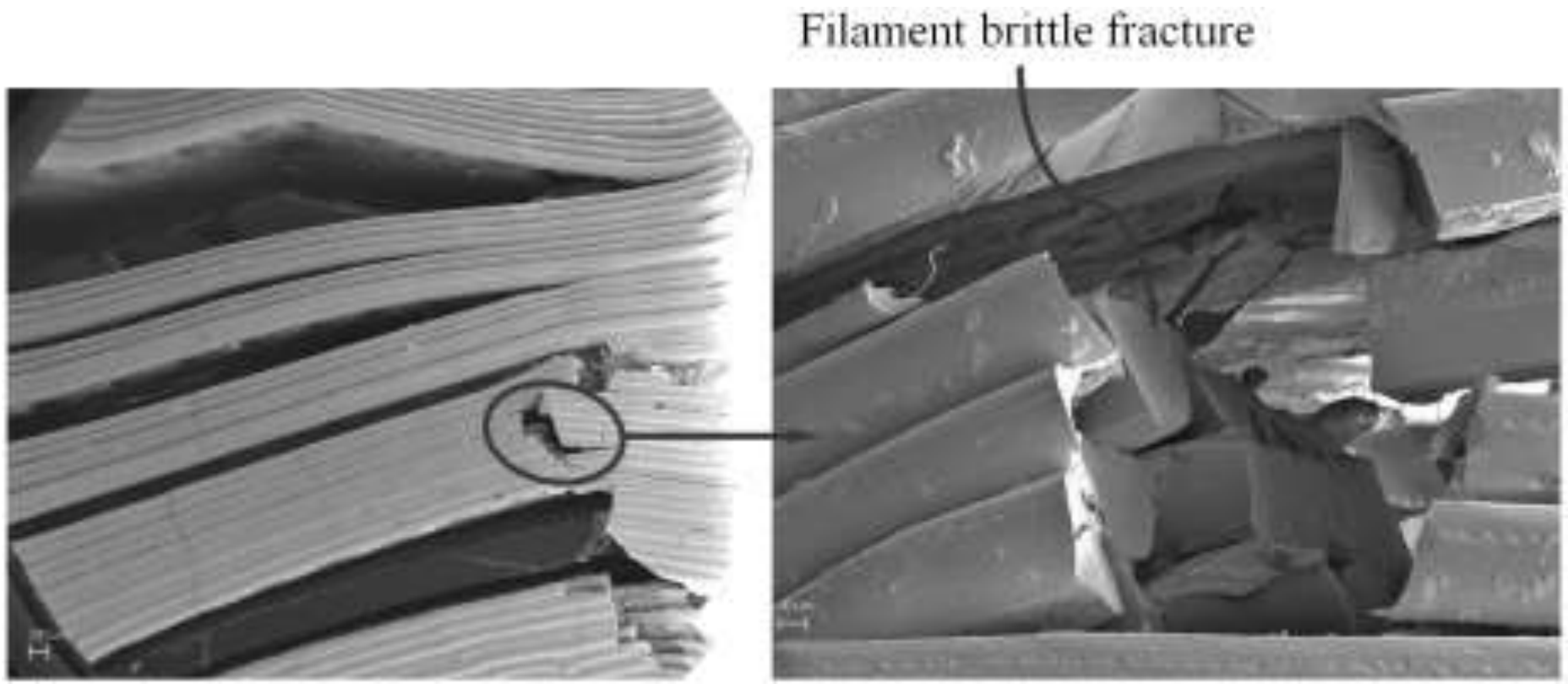

**Fig. 20** FESEM fractures images for X–axis printing direction specimens.

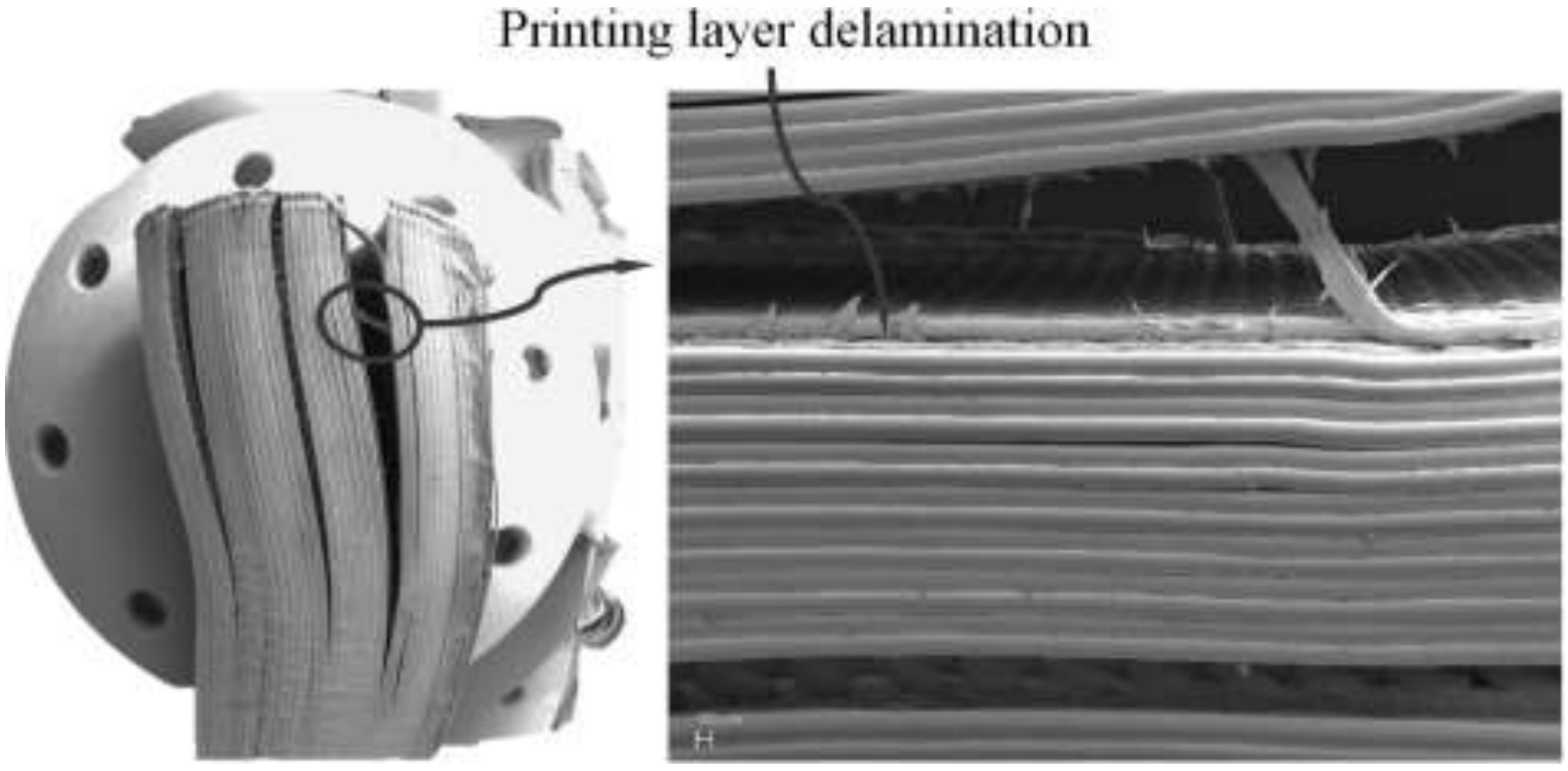

**Fig. 21** FESEM fracture images for X/Y–axis printing direction specimens

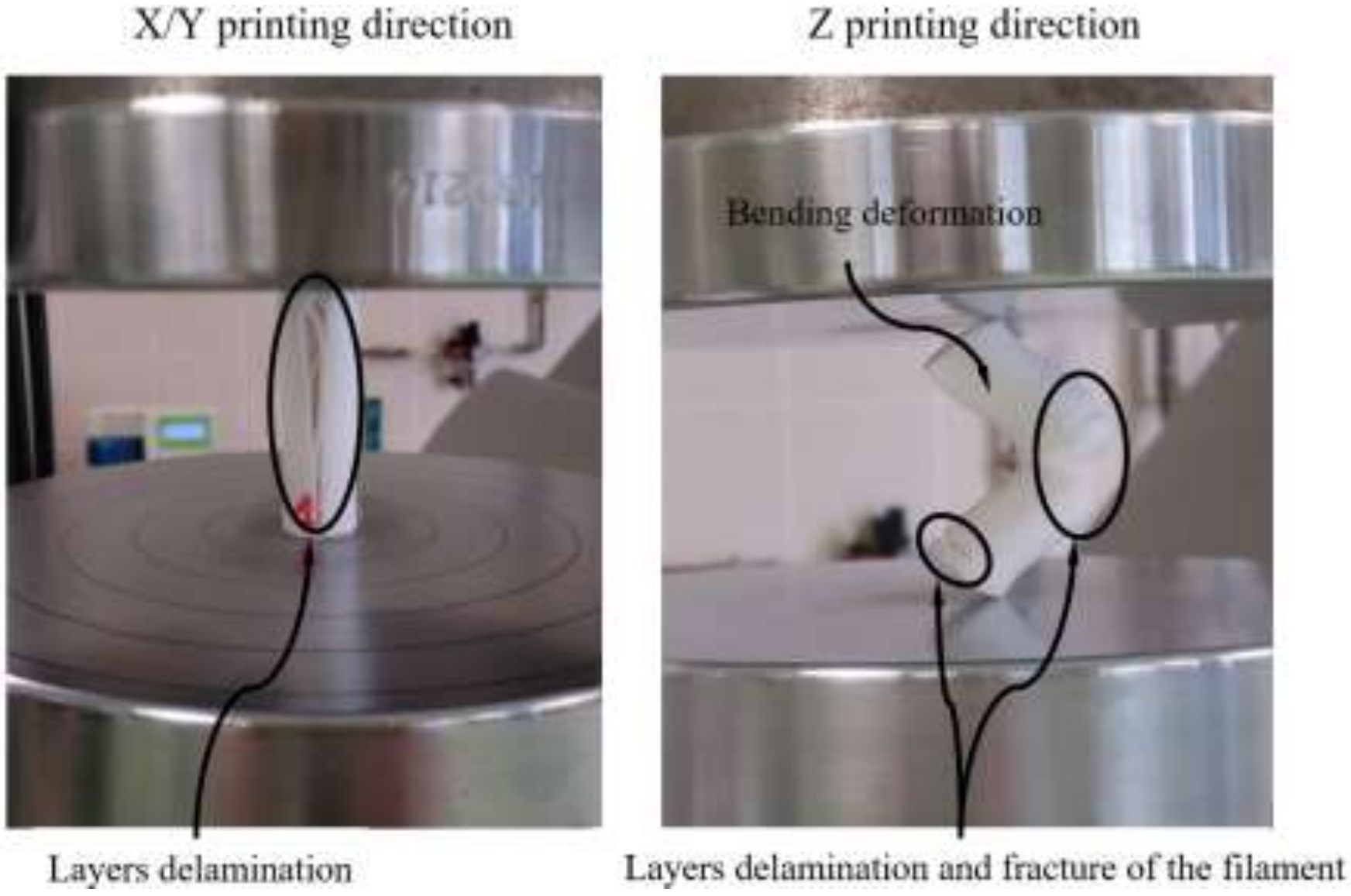

**Fig. 22** Structural collapse process of the cylindrical specimens during the experimental tests

## 6. Conclusions

The paper presents an innovative methodology based on the use of a new geometric predictive algorithm created by the authors capable of obtaining the elastic modulus of plastic parts manufactured with MEX technology as well as its behavior in the elastic zone under compression loads only taking as input parameter the material isotropic elastic modulus and the manufacturing values of the MEX process. The innovative algorithm developed computes the layer stiffness by enumerating the voids within the raised sections formed at the intersections of various virtual planes spaced according to the component's layer height.

A set of experimental tests following the standard ASTM D695–15 for a total of 42 specimens of four different 3D printing infill patterns, (concentric, Zig–Zag, lines 0º/90º, and lines –45º/45º), and three printing directions (Z–axis and X/Y–axis) have been performed. Likewise, and with the aim of validating the proposed methodology, a case study of variable topology has been experimentally tested. The comparison between predictive and experimental results shows errors between 0.2% and 4.3%, finding the X/Y direction that the best elastic and structural behavior presented because most part of the material filament is longitudinally aligned with the direction of application of the load scenario. Likewise, the 3D concentric infill pattern is the configuration that presents the best elastic and structural behavior. In this way, it is demonstrated that the predictive model can be used to obtain the elastic behavior of the plastic material manufactured with MEX process and under compression loads since the relative error does not exceed in any case the 5 % of the experimental results. In order to compare the results of the model against the results of the numerical simulation software a set of numerical analyzes using the CAE software have been proposed. The numerical results show a larger error varying between 1.1% and 13.5%. The main reason is that the numerical simulation does not evaluate the volume of holes that the geometry presents. To validate the predictive algorithm with another plastic material, the methodology outlined in this manuscript was applied and confirmed through a comparative case study, previously conducted by the same authors. Analysis reveals that both the experimental and predictive outcomes obtained through the proposed algorithm demonstrate equivalence, with a confidence level or relative error of only 0.4%

The predictive algorithm holds a key advantage in its universality, making it applicable to any plastic material and geometry produced through the MEX additive manufacturing process. This novel predictive model significantly enhances the accuracy of results compared to existing simulation software, as the latter currently lacks the capability to consider the real porosity of the component. The new algorithm improves the state of the art since, in addition to being able to obtain experimental values at no cost, is valid for most polymer plastic material, for its mechanical behavior and elastic - linear regime, and complex parts with geometric non-linearities. With the authors' algorithm, it is possible to approach the design of a complex component subjected to compressive uniaxial loads safely without the need for mechanical analysis software or expensive experimental validations. This model can help many companies that do not have expensive simulation software or do not have high incomes to make experimental tests for each designed part.

## Appendix A. Supporting information

This section presents the supplementary data associated with the experimental mechanical tests of pure uniaxial compression performed in the present research work, for the plastic material PETG manufactured with MEX process. As it shown, Fig. 23 and Fig. 24 describe the results obtained from the tests for each type of cylindrical specimen analyzed. As can be seen, for each 3D printing infill pattern, Concentric, Zig–Zag, Lines 0º/90º and Lines –45º/45º, and for each 3D printing direction, Z and X/Y–axis, a compressive curve stress – nominal strain has been generated.

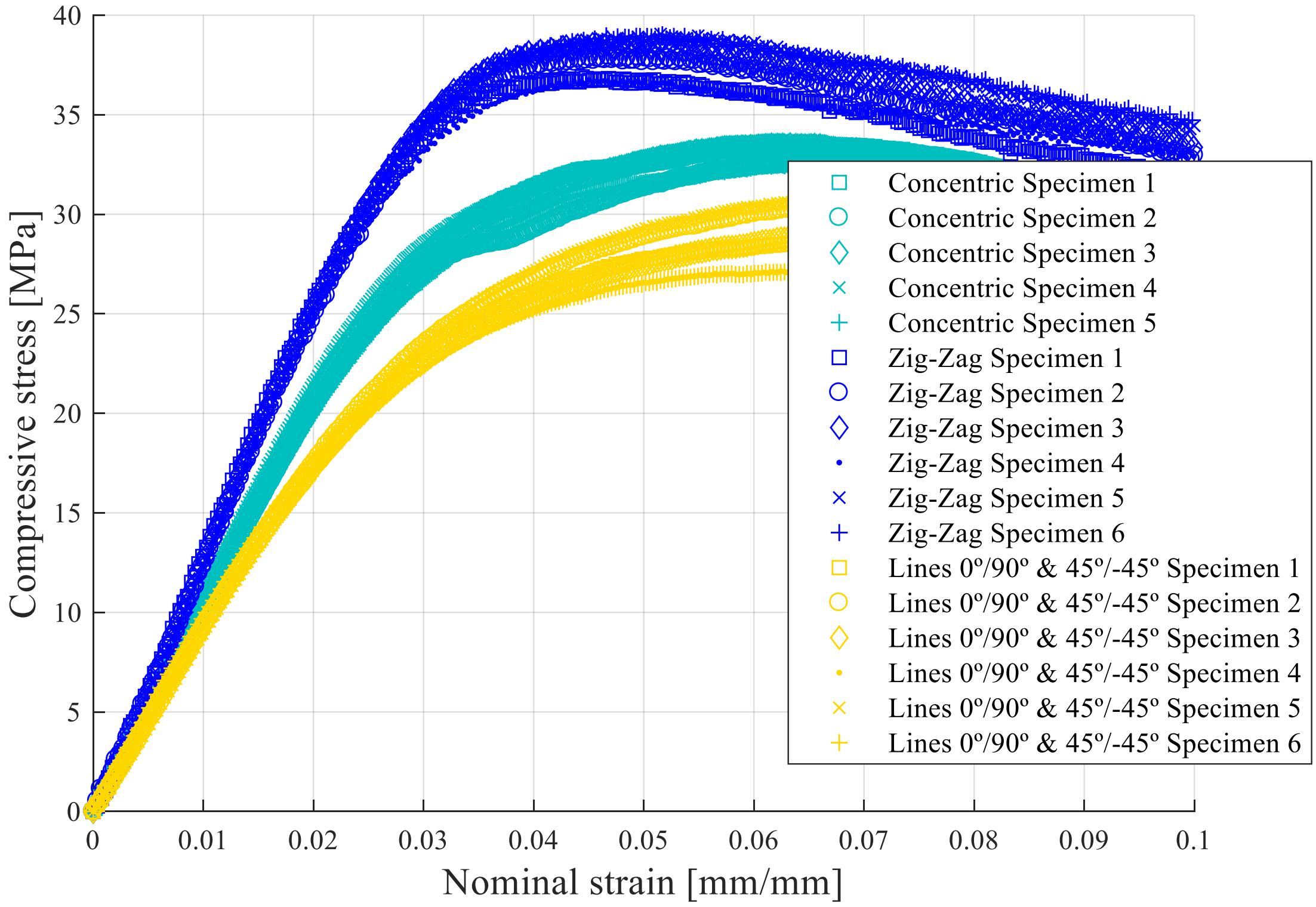

**Fig. 23** Compression experimental tests results for the Z–axis printed specimens

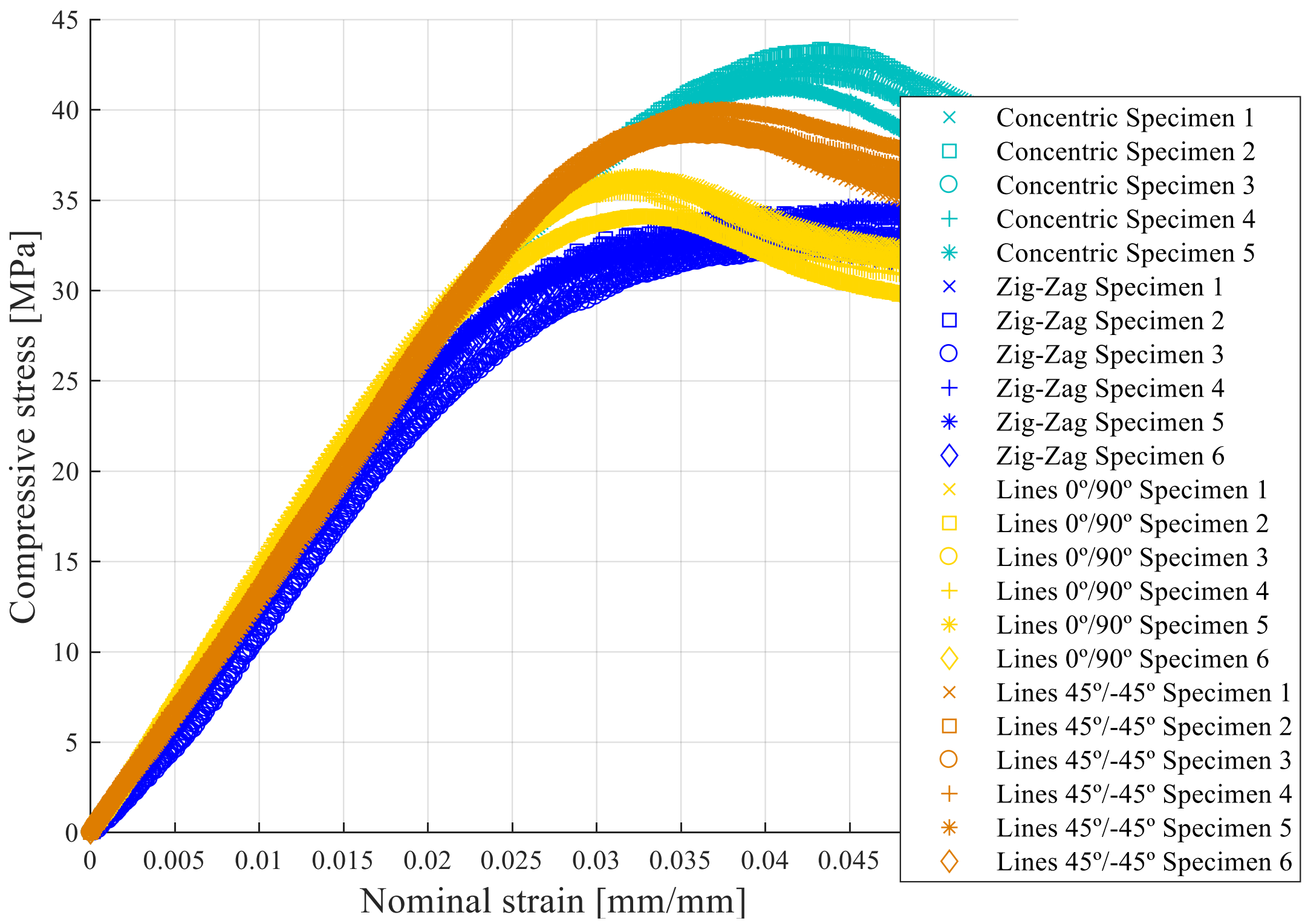

**Fig. 24** Compression experimental tests results for the X/Y–axis printed specimens

Moreover, Fig.25, Fig. 26, Table 10, and Table 11 show, for the geometry under study, the magnitude of the resulting nominal displacement obtained for a uniaxial compression force of 600 N (see Fig. 14), applied on the end of the part.

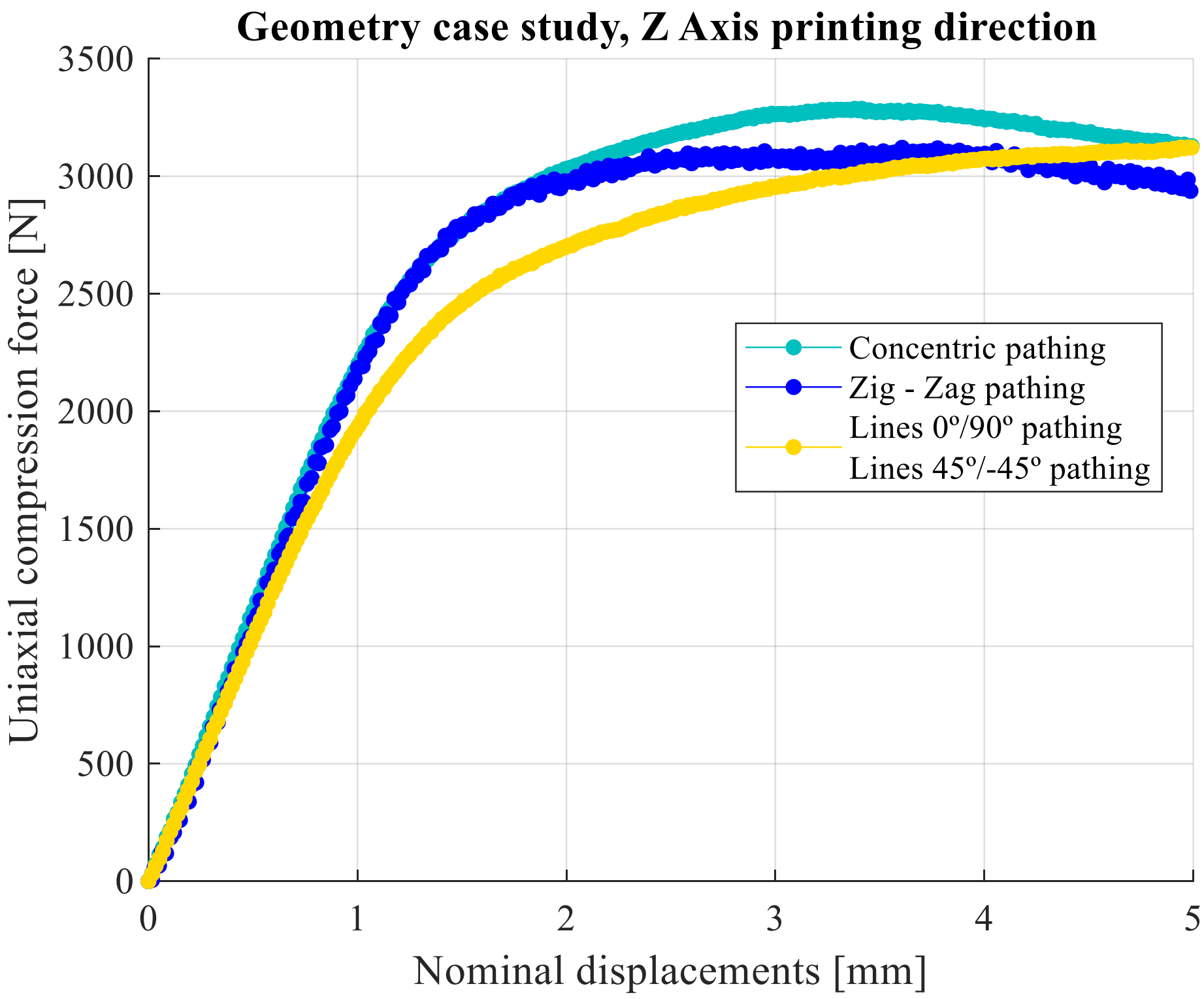

**Fig. 25** Compression experimental tests results for the Z–axis printed geometry case study

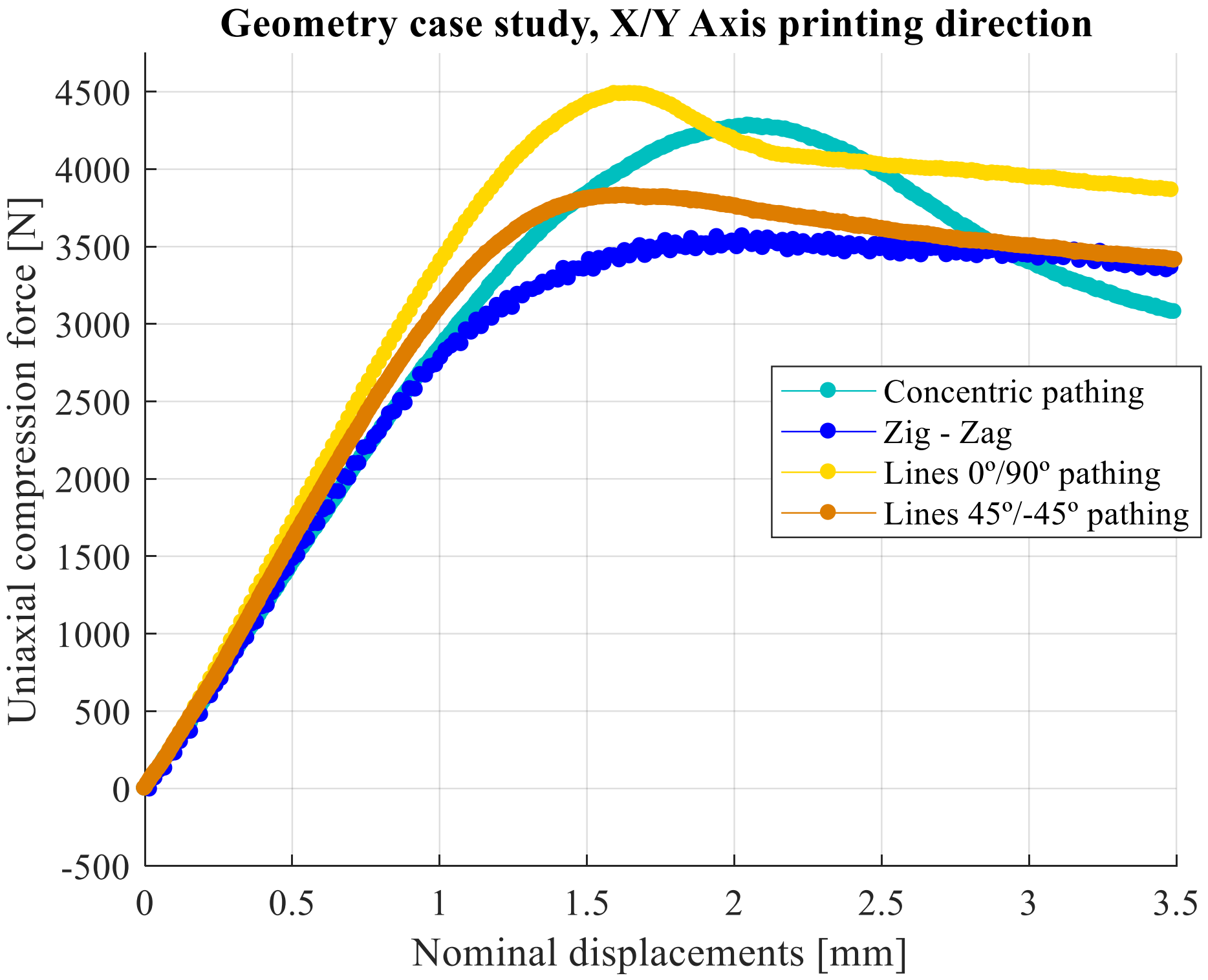

**Fig. 26** Compression experimental tests results for the X/Y–axis printed geometry case study

Finally, Fig. 27 shows the flow chart that describes the geometrical predictive algorithm defined to establish the displacement field of the geometry case study and the compressive Young's modulus of the plastic material.

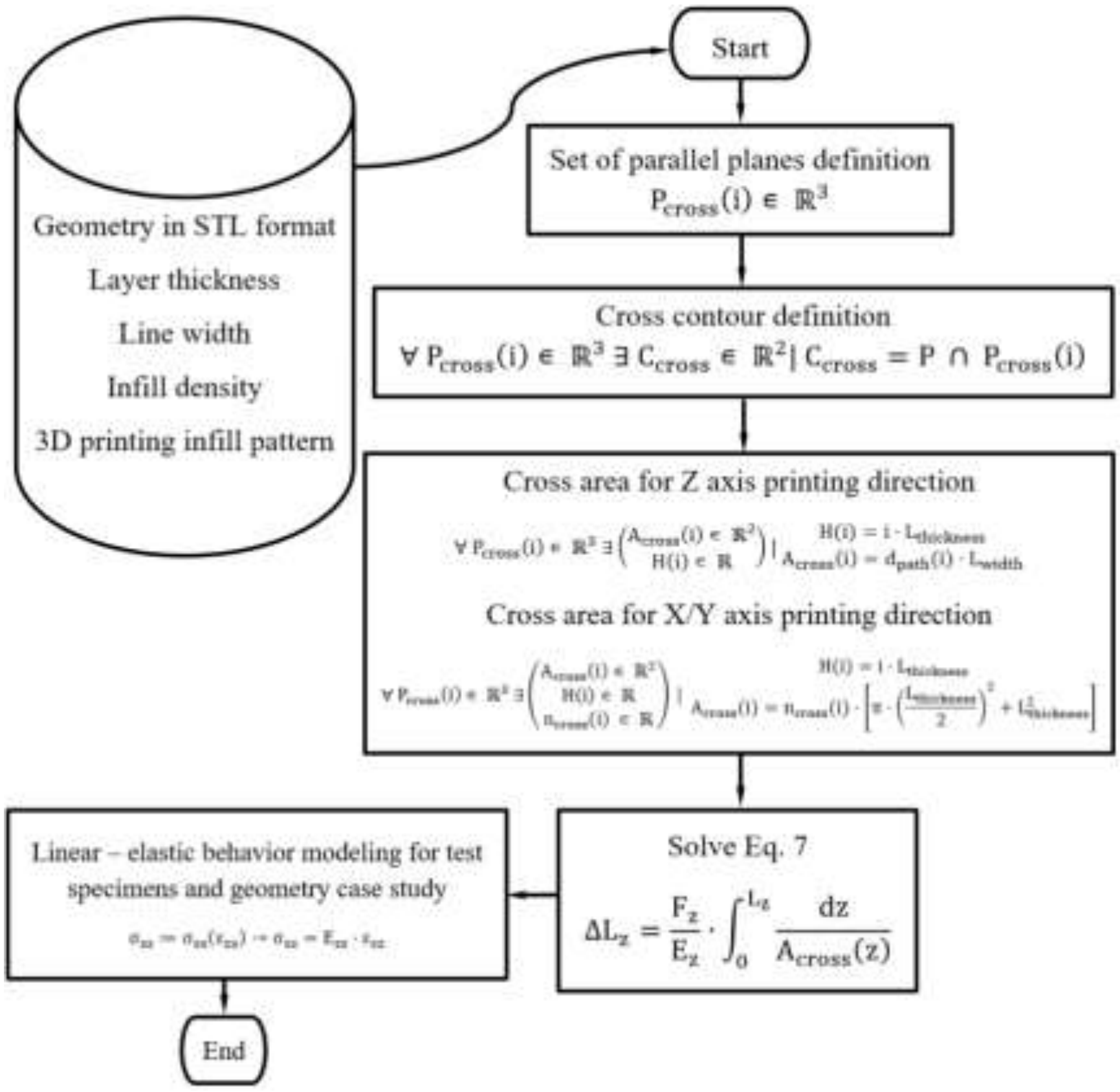

**Fig. 27** Description of the geometrical predictive algorithm defined to establish the displacement field of the geometry case study and the compressive Young's modulus of the plastic material

**Funding**

This research work was supported by the University of Jaen through the Plan de Apoyo a la Investigación 2021–2022-ACCION1a POAI 2021–2022: TIC-159

**Acknowledgments:**

The authors acknowledge the collaboration in this work with the company SmartMaterials dedicated to Industrial Design and Polymers Advanced Manufacturing

**Conflict of Interest**

The authors declare that they have no conflict of interest. Figures and Tables by authors

**Author Contributions Statement**

Investigation, J.M.M.-C., and C.M.-D.; project administration, C.M.-D.; writing—original draft, J.M.M.-C. and C.M.-D.; writing—review & editing J.M.M.-C., and C.M.-D.; funding acquisition, C.M.-D. All authors have reviewed the manuscript.